\documentclass[a4paper,12pt]{article}

\usepackage{graphicx,appendix,psfrag,amsmath,amssymb,color}

\RequirePackage[sort&compress,square,comma,numbers]{natbib}
\RequirePackage{hyperref}%
\RequirePackage{hypernat}

\allowdisplaybreaks[1]
\newcommand{\bea}{\begin{eqnarray}}
\newcommand{\eea}{\end{eqnarray}}
\newcommand{\nn}{\nonumber}
\newcommand{\be}{\begin{equation}}
\newcommand{\ee}{\end{equation}}
\newcommand{\sw}{s_{\theta_W}}
\newcommand{\cw}{c_{\theta_W}}
\newcommand{\ff}{f\hspace{-0.17cm}f}
\newcommand{\ep}{\epsilon}
\newcommand{\ord}{{\cal{O}}}
\newcommand{\as}{\alpha_s}
\newcommand{\aem}{\alpha_{em}}

\newcommand{\captionfonts}{\small}
\makeatletter  
\long\def\@makecaption#1#2{%
  \vskip\abovecaptionskip
  \sbox\@tempboxa{{\captionfonts #1: #2}}%
  \ifdim \wd\@tempboxa >\hsize
    {\captionfonts #1: #2\par}
  \else
    \hbox to\hsize{\hfil\box\@tempboxa\hfil}%
  \fi
  \vskip\belowcaptionskip}
\makeatother

\begin{document}

\begin{titlepage}

\begin{flushright}
\normalsize
MZ-TH/11-42\\
November 28, 2011
\end{flushright}

\begin{center}
\Large\bf\boldmath
{Soft-gluon resummation for slepton-pair production at hadron colliders}
\unboldmath
\end{center}

\vspace*{0.8cm}
\begin{center}

{\sc Alessandro Broggio, Matthias Neubert and Leonardo Vernazza\footnote{Alexander von Humboldt Fellow}}\\[0.7cm]
{\sl Institut f\"ur Physik (THEP)\\ 
Johannes Gutenberg-Universit\"at, D--55099 Mainz, Germany}\\[0.4cm]
\end{center}

\vspace*{0.8cm}
\begin{abstract}
We use a recent approach to soft-gluon resummation based on effective field theory to implement soft-gluon resummation at NNNLL order for slepton-pair production in SUSY extensions of the Standard Model. This approach resums large logarithmic corrections arising from the dynamical enhancement of the partonic threshold region caused by steeply falling parton luminosities. We evaluate the resummed invariant-mass distribution and total cross section for slepton-pair production at the Tevatron and LHC, matching our results onto NLO fixed-order calculations. As a byproduct, we also study virtual SUSY effects in the context of soft-gluon resummation for the Drell-Yan process.
\end{abstract}
\vfil

\end{titlepage}

\section{Introduction}

With the advent of the Large Hadron Collider (LHC), the experimental investigation of TeV-scale physics is now fully accessible. The stabilization of the electroweak scale requires new particles and interactions in the TeV range, and supersymmetry (SUSY) is one of the most compelling scenarios which achieves such a stabilization. For every Standard Model (SM) particle it introduces a corresponding superpartner, whose spin differs by 1/2 from that of the SM particle. The non-observation of these particles so far requires a (soft) breaking of SUSY, and the superpartners are expected to have masses in the TeV range. Therefore, it should be possible to observe these particles through their direct production at hadron colliders, and indirectly through their virtual contributions to SM-like processes.

In this paper we analyze the production of a scalar-lepton pair at hadron colliders \cite{Dawson:1983fw,Chiappetta:1985ku,delAguila:1990yw,Baer:1993ew}, which we consider in parallel with the classic Drell-Yan production of a lepton pair \cite{Drell:1970wh} in SUSY. These two processes can be considered together because they arise from the same hard-scattering interaction: the annihilation of a quark-antiquark pair into a virtual photon or $Z$ boson, which then decays into a slepton or a lepton pair, respectively. The effect of strongly-interacting SUSY particles enters in the hard-scattering interaction of both processes only at the one-loop level through the virtual exchange of squarks and gluinos. Both processes are interesting and play an important role: Drell-Yan production can be considered as a prototype for other collider processes and, among other things, its cross section as a function of the invariant mass of the lepton pair can be used to search for new heavy resonances. Sleptons are expected to be among the lightest SUSY particles, which means that in many scenarios they decay directly into the corresponding SM partners and the stable lightest SUSY particle, giving rise to simple signatures such as a pair of energetic leptons plus missing energy.

Both processes have been studied extensively in the past. The calculation of the cross section and rapidity distribution at the next-to-leading order (NLO) in $\as$ for the Drell-Yan process in the SM has been accomplished long ago \cite{Altarelli:1979ub}, while the corresponding results at the next-to-next-to-leading order (NNLO) were obtained more recently \cite{Hamberg:1990np,Harlander:2002wh,Anastasiou:2003yy,Anastasiou:2003ds,Melnikov:2006di,Melnikov:2006kv}. A study of SUSY QCD and electroweak corrections at NLO was performed in \cite{Dittmaier:2009cr}. Results for the total cross section for slepton-pair production at NLO in $\as$ were obtained in \cite{Baer:1997nh,Beenakker:1999xh,Djouadi:1999ht}. The main uncertainties in the theoretical predictions arise from the imperfect knowledge of the parton distribution functions (PDFs) and from the truncation of the perturbative expansion, which introduces a dependence on the unphysical factorization and renormalization scales. The two sources of errors are of comparable size. In particular, the uncertainty due to scale variations  is smaller than in similar production processes involving colored particles in the final states, like top-quark pair production. This is because, at the partonic tree level, (s)lepton-pair production is a purely electroweak process, and therefore at leading order the uncertainty arise only from the variation of the factorization scale of the PDFs. The uncertainty from the renormalization scale starts at order $\as$ and is therefore suppressed. A reduction of the scale uncertainties is nevertheless  desirable, because having a small error on the cross section and the differential distributions allows one to extract interesting information, such as the slepton masses, with better precision.

Reducing the scale uncertainty by computing higher-order terms in the pertubative expansion is a task which soon becomes prohibitive. A good alternative for improving the theoretical predictions is to resort to soft-gluon resummation methods, which allow one to take into account the dominant contributions of the higher-order terms \cite{Sterman:1986aj,Catani:1989ne}. These contributions arise from large Sudakov logarithms, which originate as a left-over from the cancellation of virtual and real soft divergences in a kinematical configuration where the invariant mass $M$ of the (s)lepton pair is close to the partonic center-of-mass energy, and hence there is little energy left for additional real radiation. These logarithms must be resummed to all orders to improve the convergence of the perturbative expansion. The resummation was first achieved at next-to-leading-logarithmic (NLL) order for the Drell-Yan invariant-mass distribution in \cite{Sterman:1986aj,Catani:1989ne}, based on a method involving the solution of certain evolution equations in Mellin moment space. It was later extended to the rapidity distribution \cite{Sterman:2000pt}, and to N$^3$LL order in \cite{Ravindran:2006bu,Ravindran:2007sv}. In the case of slepton-pair production, the resummation was performed at NLL level for the invariant-mass distribution and total cross section \cite{Bozzi:2007qr}. Recently, it was shown that Sudakov logarithms can be resummed directly in momentum space using methods of soft-collinear effective theory (SCET) \cite{Becher:2006nr}. This approach allows for straightforward operator definitions of the various terms which build up the cross section, and it avoids the appearance of Landau poles in the resummation procedure. The method was used to perform the threshold resummation for the Drell-Yan production up to N$^3$LL order in \cite{Becher:2007ty}, and then extended to Higgs production and top-quark pair production \cite{Ahrens:2008qu,Ahrens:2008nc,Ahrens:2009uz,Ahrens:2010zv}.

The aim of our paper is to extend previous analyses in various directions. First, we extend the results of \cite{Becher:2007ty} for the Drell-Yan production of lepton pairs by including the contribution from a virtual $Z$ boson as well as SUSY QCD corrections at order $\as$. In the case of slepton-pair production, we extend the results  of \cite{Bozzi:2007qr} by performing the threshold resummation at N$^{3}$LL order. While this has a minor effect on the total correction to the differential distributions and cross section, it may be relevant for the theoretical uncertainty estimate. Moreover, since our resummation is performed by means of SCET methods, which constitute an alternative to the Mellin-moment technique used in \cite{Bozzi:2007qr}, we have an independent check of the results in literature. We present results for the differential and total cross sections for the Tevatron and the LHC with a center-of-mass energy of 7 and 14\,TeV. 

In Section~\ref{kinfact} we recall the basic formulas for the differential distributions and define the kinematics of the threshold region. We then review in Section~\ref{factscet} the factorization of the cross section and the resummation of the threshold logarithms in SCET, presenting the result specific to the SUSY case. Section~\ref{pheno} is devoted to a comprehensive phenomenological analysis. We estimate the relevance of SUSY QCD corrections and the impact of soft-gulon resummation on the invariant-mass distribution and the total cross section for slepton-pair production. We also discuss the uncertainties due to scale variations and the errors on the PDFs. Our conclusions are given in Section~\ref{concl}.

\section{Kinematics and factorization at threshold}
\label{kinfact}

We consider the production of a (s)lepton pair with invariant mass $M$ in hadron-hadron collisions, at center-of-mass energy $\sqrt{s}$. The process involves the reaction $N_1(P_1)+N_2(P_2)\to\gamma^*/Z^{0*}+X$, where $X$ represents an inclusive hadronic final state, followed by $\gamma^*/Z^{0*}\to\tilde l^-(p_3)+\tilde l^+(p_4)$ or $l^-(p_3)+l^+(p_4)$. We start by focusing on the double-differential cross section in the invariant mass $M^2=q^2$ and rapidity $Y=\frac{1}{2}\ln \frac{q^0+q^3}{q^0-q^3}$ of the (s)lepton pair in the center-of-mass frame, where $q=p_3+p_4$. This cross section can be calculated up to power corrections in perturbative QCD and expressed in terms of convolutions of short-distance partonic cross sections with PDFs:
\be\label{a1}
   \frac{d^2\sigma}{dM^2dY}
   = \sigma_0 \sum_{ij} \int dx_1 dx_2\,\widetilde{C}_{ij}(x_1,x_2,s,M,\mu_f)\,
    f_{i/N_1}(x_1,\mu_f)\,f_{j/N_2}(x_2,\mu_f) \,,
\ee
where $\mu_f$ is the factorization scale, $f_{i/N}(x,\mu_f)$ gives the probability of finding a parton $i$ with longitudinal momentum fraction $x$ inside the hadron $N$, and
\be\label{a2}
   \sigma_0 = \frac{4\pi\aem^2}{3N_c M^2 s} \quad \mbox{for $l^- l^+$} \,,\qquad
   \sigma_0 = \frac{\pi\aem^2\beta_{\tilde l}^3}{3N_c M^2 s} \quad 
    \mbox{for $\tilde l^-\tilde l^+$} \,,
\ee
with $\beta_{\tilde l}=\sqrt{1-4m_{\tilde l}^2/M^2}$ denoting the 3-velocity of the slepton in the $\tilde l^-\tilde l^+$ rest frame. The hard-scattering kernels $\widetilde{C}_{ij}$ are related to the partonic cross sections and can be calculated as power series in $\as$. At leading order ($\sim\as^0$) the sum involves only the channels $(ij)=(q \bar{q}), (\bar{q} q)$, with $p_1= x_1 P_1$, $p_2= x_2 P_2$. At NLO ($\sim\as$) one has to take into account $(ij)= (q\bar q), (\bar q q), (qg), (gq), (\bar q g), (g\bar q)$. Here we are interested in the evaluation of higher-order radiative corrections near threshold, for which it is useful to define the quantities
\be
   \tau = \frac{M^2}{s} \,, \qquad 
   z = \frac{M^2}{\hat s} = \frac{\tau}{x_1 x_2} \,,
\ee
where $\hat s= x_1 x_2 s$ is the partonic center-of-mass energy squared. The partonic threshold region is defined by the limit $z\to 1$, in which the dynamics of the process is greatly simplified. Since the partonic center-of-mass energy is just sufficient to create the (s)lepton pair, there is no phase space available for the emission of additional energetic partons. The cross section is dominated by the terms which are singular in the $z\to 1$ limit, which correspond to the virtual corrections and the real emission of soft gluons. Such terms only arise for the $(q\bar q)$ and $(\bar q q)$ channels.

We define the couplings of the (s)fermions to gauge boson $i$ following the notation of \cite{Becher:2010tm}. The relevant chiral currents
\bea\label{a4}
   J_{f,i}^\mu &=& \sum_f \left( g_L^{f,i} \bar f\gamma^\mu P_L f
    + g_R^{f,i} \bar f\gamma^\mu P_R f \right) , \nn\\
   J_{\tilde f,i}^\mu &=& \sum_{\tilde f} 
    \left( g_L^{\tilde f,i} \tilde f_L^* \overleftrightarrow{\partial}^{\!\mu} \tilde f_L
    + g_R^{\tilde f,i} \tilde f_R^* \overleftrightarrow{\partial}^{\!\mu} \tilde f_R \right) ,
\eea
with $P_{L/R}=\frac12\,(1\mp\gamma_5)$, involve the couplings
\be
\begin{array}{ll}
  g_{L}^{q,\gamma} = g_{L}^{\tilde{q},\gamma} = e_q\,, \qquad\qquad &
  g_{R}^{q,\gamma} = g_{R}^{\tilde{q},\gamma} = e_q\,, \\
  g_{L}^{q,Z} = g_{L}^{\tilde{q},Z} = \frac{-1+\frac{4}{3}\sw^2}{2\sw\cw}\,, \qquad\qquad &
  g_{R}^{q,Z} = g_{R}^{\tilde{q},Z} = \frac{\frac{4}{3}\sw^2}{2\sw\cw}\,, \\
  g_{L}^{l,\gamma} = g_{L}^{\tilde{l},\gamma} = 1\,, \qquad\qquad &
  g_{R}^{l,\gamma} = g_{R}^{\tilde{l},\gamma} = 1\,, \\
  g_{L}^{l,Z} = g_{L}^{\tilde{l},Z} = \frac{1-2\sw^2}{2\sw\cw}\,, \qquad\qquad &
  g_{R}^{l,Z} = g_{R}^{\tilde{l},Z} = \frac{-2\sw^2}{2\sw\cw}\,, \\
  g^{\tilde \tau_1,Z} = \frac{\cos\theta_{\tilde \tau}-2\sw^2}{2\sw\cw}\,, \qquad\qquad &
  g^{\tilde \tau_2,Z} = \frac{\sin\theta_{\tilde \tau}-2\sw^2}{2\sw\cw}\,.
\end{array}
\ee
Here $\sw=\sin\theta_W$, $\cw=\cos\theta_W$, where $\theta_W$ is the electroweak mixing angle. We consider the possibility of mixing between the third-generation sleptons, introducing the mass eigenstates $\tilde\tau_1$, $\tilde\tau_2$ and the corresponding mixing angle $\theta_{\tilde\tau}$. 

The leading contributions to the double-differential cross section arising in the near partonic threshold $z\to 1$ can be written as \cite{Becher:2007ty}
\bea\label{a6}
   \frac{d^2\sigma^{\rm thresh}}{dM^2dY}
   &=& \sigma_0 \sum_q\,h_q^{(l,\tilde l)} \int\frac{dz}{z}\,
    C(z,M,m_{\tilde q},m_{\tilde g},\mu_f) \\
   &&\hspace{-2.0cm}\times \bigg[ 
    \frac{f_{q/N_1}(\sqrt{\tau}e^Y,\mu_f)\,f_{\bar q/N_2}(\frac{\sqrt{\tau}}{z}e^{-Y},\mu_f) 
    + f_{q/N_1}(\frac{\sqrt{\tau}}{z}e^Y,\mu_f)\,f_{\bar q/N_2}(\sqrt{\tau}e^{-Y},\mu_f)}{2}
    + (q\leftrightarrow\bar q) \bigg] \,. \nn
\eea
The coefficients $h_q^{(l,\tilde l)}$ take into account the photon, $Z$ boson, and $\gamma$-$Z$ interference contributions. For lepton-pair production they read
\be
   h_q^{(l)} = \left[ e_q^2
    - \frac12\,\frac{e_q(g_L^{q,Z}+g_R^{q,Z})(g_L^{l,Z}+g_R^{l,Z})}{1-m_Z^2/M^2}
    + \frac14\,\frac{({g_L^{q,Z}}^2+{g_R^{q,Z}}^2)({g_L^{l,Z}}^2+{g_R^{l,Z}}^2)}%
                    {\left(1-m_Z^2/M^2\right)^2} \right] ,
\ee
while for the production of a slepton pair of type $\tilde l$
\be
   h_q^{(\tilde l)} = \left[ e_q^2
    - \frac{e_q(g_L^q+g_R^q)g^{\tilde l,Z}}{1-m_Z^2/M^2}
    + \frac12\,\frac{({g_L^q}^2+{g_R^q}^2){g^{\tilde l,Z}}^2}{\left(1-m_Z^2/M^2\right)^2}
    \right] .
\ee
The hard-scattering kernel $C(z,M,m_{\tilde q},m_{\tilde g},\mu_f)$ in (\ref{a6}) contains both SM and SUSY QCD corrections. The SM QCD corrections are known to two-loop order \cite{Anastasiou:2003yy}. The SUSY corrections arising at NLO the latter are given by a vertex diagram with a gluino-squark-squark loop plus external-leg corrections. They have been calculated in \cite{Beenakker:1999xh,Bozzi:2007qr,Bozzi:2007tea,Dittmaier:2009cr}. We have recomputed these contributions and find agreement with results in the literature. Integrating over the rapidity, one obtains from (\ref{a6}) the single-differential cross section
\be\label{a9}
   \frac{d\sigma^{\rm thresh}}{dM^2}
   = \sigma_0 \sum_q\,h_q^{(l,\tilde l)} \int_{\tau}^1\frac{dz}{z}\,
    C(z,M,m_{\tilde q},m_{\tilde g},\mu_f)\,\ff(\tau/z,\mu_f) \,,
\ee
where
\be
   \ff(y,\mu_f) = \int_y^1\!\frac{dx}{x} \left[ f_{q/N_1}(x,\mu_f)\,f_{\bar q/N_2}(y/x,\mu_f)
    + (q\leftrightarrow\bar q) \right]
\ee
defines the parton luminosity function.

\section{Factorization in SCET}
\label{factscet}

The hard-scattering kernel $C(z,M,m_{\tilde q},m_{\tilde g},\mu_f)$ depends on the invariant mass $M$ and on the masses of squarks and gluinos, $m_{\tilde q}$ and $m_{\tilde g}$. We will assume in our analysis that these three scales are of similar order. For $z\to 1$ one can then distinguish two well separated mass scales, the ``hard'' scale $\mu_h\sim M$ and the ``soft'' scale $\mu_s\sim M(1-z)/\sqrt{z}=\sqrt{\hat{s}}(1-z)$, which corresponds to the energy of the emitted soft gluons. The presence of these two scales is reflected in the factorization of the coefficient $C(z,M,m_{\tilde q},m_{\tilde g},\mu_f)$ into a hard and a soft function,
\be
   C(z,M,m_{\tilde q},m_{\tilde g},\mu_f) 
   = H(M,m_{\tilde q},m_{\tilde g},\mu_f)\,S(\sqrt{\hat s}(1-z),\mu_f) \,.
\ee
Choosing the factorization scale $\mu_f$ in (\ref{a6}) close to $\mu_h$ or $\mu_s$ unavoidably causes the appearance of large logarithms in one of the two factors. Threshold resummation addresses the problem of resumming these large logarithms to all orders in perturbation theory.

Recently, the factorization formula (\ref{a9}) has been obtained in the context of SCET \cite{Becher:2007ty}, which is the approach we follow in this paper. It is not necessary to repeat the derivation here. It suffices to remember that, considering for simplicity the Drell-Yan lepton-pair production through the exchange of a virtual photon with total momentum $q^\mu$, one starts with the expression of the cross section written in terms of the currents defined in (\ref{a4}),
\be
   d\sigma = \frac{4\pi\aem^2}{3sq^2}\frac{d^4q}{(2\pi^4)} \int d^4x\,e^{-iq\cdot x}
    \langle N_1(p_1)N_2(p_2)|(-g_{\mu\nu}) J_{q,\gamma}^{\mu\dag}(x)\,
    J_{q,\gamma}^\nu(0)|N_1(p_1)N_2(p_2)\rangle \,.
\ee
The factorization theorem is derived by matching the product of currents onto operators in SCET and decoupling soft and collinear interactions by means of field redefinitions. In the first step, the current $J_{q,\gamma}^\mu$ is matched onto
\be\label{a13}
   J_{q,\gamma}^\mu \rightarrow C_V(-q^2-i\ep,m_{\tilde q},m_{\tilde g},\mu)
    \sum_q \left( g_L^q \bar\chi_{\overline{hc}}S^{\dag}_{\bar n} \gamma^{\mu} P_L 
    S_n \chi_{hc}
    + g_R^q \bar\chi_{\overline{hc}}S^{\dag}_{\bar n} \gamma^{\mu} P_R S_n \chi_{hc} \right) ,
\ee
where the effective fields $\chi_{hc}=W_{hc}^{\dag}\xi_{hc}$ and $\chi_{\overline{hc}}=W_{\overline{hc}}^{\dag}\xi_{\overline{hc}}$ are the gauge-invariant combinations of (anti-)hard-collinear quark fields and Wilson lines in SCET. The matching coefficient $C_V$ depends on the time-like, hard momentum transfer $Q^2=-q^2=-M^2$. It is given by the on-shell massless quark form factor \cite{Becher:2006nr}, which in the present case must be calculated including SUSY QCD corrections. Given that the cross section for (s)lepton-pair production involves the current squared, it follows that
\be
   H(M,m_{\tilde q},m_{\tilde g},\mu_h) 
   = \left| C_V(-M^2-i\ep,m_{\tilde q},m_{\tilde g},\mu_h) \right|^2 .
\ee
The soft function arises in a second matching step, in which the soft fields contained in the Wilson lines $S_{n}$ and $S_{\bar n}$ in (\ref{a13}) are integrated out. Since it is insensitive to short-distance physics, the soft function is the same for Drell-Yan and slepton-pair production. It is defined in terms of the vacuum matrix element of a correlator formed by the product of the soft Wilson lines contained in the two currents \cite{Becher:2007ty}. 

The resummation of threshold logarithms can be achieved directly in momentum space by solving the renormalization-group (RG) equations for the hard and soft functions. In this way, one obtains for the hard-scattering kernel the compact expression 
\bea\label{a14}
   C(z,M,m_{\tilde q},m_{\tilde g},\mu_f) 
   &=& \left| C_V(-M^2,m_{\tilde q},m_{\tilde g},\mu_h) \right|^2
    U(M^2,\mu_h,\mu_s,\mu_f) \nn\\
   &&\times\frac{z^{-\eta}}{(1-z)^{1-2\eta}}\,
    \tilde s_{\rm DY}\bigg(\ln \frac{M^2(1-z)^2}{\mu_s^2 z}+\partial_{\eta},\mu_s\bigg)\,
    \frac{e^{-2\gamma_E\eta}}{\Gamma(2\eta)} \,,
\eea
where the definitions of the evolution function $U$, the coefficient $\eta$ and the Laplace-transformed soft function $\tilde s_{\rm DY}$ can be found in \cite{Becher:2007ty}. The above expression accomplishes the resummation of the leading terms for $z\to 1$ to all orders of perturbation theory.

In the SM, the expression for $C_V$ has recently been derived up to three-loop order \cite{Gehrmann:2005pd,Gehrmann:2010ue}. Including the additional virtual corrections of SUSY particles, the perturbative expansion up to order $\ord(\as^2)$ can be written as
\bea\label{a15}
   C_V(-M^2,m_{\tilde q},m_{\tilde g},\mu_h) 
   &=& 1 + \frac{\as}{4\pi} \left[ c_V^{(1)}(-M^2,\mu_h)
    + c_{V,\rm SUSY}^{(1)}(-M^2,m^2_{\tilde{q}},m^2_{\tilde{g}}) \right] \nn\\
   &&\mbox{}+ \left( \frac{\as}{4\pi} \right)^2
    \left[ c_V^{(2)}(-M^2,\mu_h)
    + c_{V,\rm SUSY}^{(2)}(-M^2,m^2_{\tilde{q}},m^2_{\tilde{g}},\mu_{h}) \right] , \qquad
\eea
where $c^{(1)}_V$ and $c^{(2)}_V$ include the one- and two-loop QCD corrections present in the SM, while $c_{V,\rm SUSY}^{(1)}$ and $c_{V,\rm SUSY}^{(2)}$ represent the additional QCD corrections arising in its SUSY extension. In the following we will neglect the two-loop SUSY contribution, which we assume to be negligible since already $c_{V,\rm SUSY}^{(1)}$ will turn out to be very small. The explicit expressions for $c^{(1)}_V$ and $c^{(2)}_V$ can be found in \cite{Becher:2007ty}, while
\bea\label{a16}
   c_{V,\rm SUSY}^{(1)}
   &=& C_F\,\Bigg\{ \frac{5}{2} - \frac{m_{\tilde g}^2}{m_{\tilde g}^2-m_{\tilde q}^2}
    + \frac{2(m_{\tilde g}^2-m_{\tilde q}^2)}{M^2} 
    + \bigg[ \frac{m_{\tilde g}^4}{\left(m_{\tilde g}^2-m_{\tilde q}^2\right)^2}
    + \frac{2m_{\tilde g}^2}{M^2} \bigg]\,\ln\frac{m_{\tilde g}^2}{m_{\tilde q}^2} \\
   &&\mbox{}- \left[ 1 + \frac{2(m_{\tilde g}^2-m_{\tilde q}^2)}{M^2} \right]
    f_B(M^2,m_{\tilde q}^2) 
    + 2 \bigg[ \frac{m_{\tilde g}^2}{M^2} 
    + \frac{\left(m_{\tilde g}^2-m_{\tilde q}^2\right)^2}{M^4} \bigg]\,
    f_{C}(M^2,m_{\tilde q}^2,m_{\tilde g}^2) \Bigg\} . \nn
\eea
For simplicity we assume degenerate squark masses for $\tilde q_{L,R}$ with $q=u,d,s,c,b$. The loop functions $f_B$ and $f_C$ are provided in the Appendix. The evolution equation for $C_V$ in (\ref{a15}) is the same as for the corresponding coefficient in the SM. As a result, $c_{V,\rm SUSY}^{(1)}$ does not depend on the renormalization scale. Note that the Wilson coefficient $C_V$ is the same for all currents in (\ref{a4}).

\section{Systematic studies and phenomenology}
\label{pheno}

We now present a detailed numerical analysis of the invariant-mass distribution and total cross section for slepton-pair production at the Tevatron and LHC. As a byproduct, we will also study the rapidity distribution for the Drell-Yan production of lepton pairs. Our goal is to estimate the impact of soft-gluon resummation and the relevance of SUSY contributions to these observables. To this end, we will either focus on the physical cross sections directly or consider the $K$-factor defined as
\be
   \frac{d\sigma}{dM^2} 
   = K(M^2,m^2_{\tilde{q}},m^2_{\tilde{g}},\tau)\,\frac{d\sigma}{dM^2}\bigg|_{\rm LO} \,.
\ee
The theoretical predictions depend on various input parameters, whose numerical values are $\aem(M_Z)=1/128$, $\sin\theta_W=0.23143$, $M_Z=91.188$\,GeV, and $\Gamma_Z=2.4952$\,GeV. Our assumptions for the masses of SUSY particles will be discussed in more detail below. For the systematic analyses in Sections~\ref{subsec:scales} and \ref{subsec:susy} we use the PDF set MSTW2008NNLO \cite{Martin:2009iq,Martin:2009bu} and $\as(M_Z)=0.117$ with three-loop running in the $\overline{\rm MS}$ scheme. Using a fixed set of PDFs helps to illustrate more clearly the behavior of the perturbative expansion of the hard-scattering kernel in higher orders of perturbation theory. 

For appropriate choices of the hard and soft scales $\mu_h$ and $\mu_s$, the expressions for the hard-scattering kernel given in (\ref{a14}) resums the leading singular contributions to the partonic cross sections in the limit where the partonic center-of-mass energy $\sqrt{\hat s}$ is close to the invariant mass $M$ of the (s)lepton pair. In fixed-order perturbation theory, the leading terms correspond to plus distributions of the form
\be
   \bigg[ \frac{1}{1-z}\,\ln\frac{M^2(1-z)^2}{\mu_f^2\,z} \bigg]_+ .
\ee
For Drell-Yan type processes, it is well known that, after the hard-scattering kernels are convoluted with the parton luminosities, these terms provide the dominant contributions to the perturbative series for the cross sections, typically accounting for more than 90\% of the one- and two-loop corrections (see e.g.\ \cite{Becher:2007ty,Ahrens:2008nc,Ahrens:2010zv}). In realistic cases where the ratio $\tau=M^2/s$ is not very close to 1, the dominance of the region $z\lesssim 1$ in the calculation of the cross section arises dynamically, due to the strong fall-off of the parton luminosities. Below we will perform the resummation of the leading terms at different orders in RG-improved perturbation theory. At NNLL order, one evaluates (\ref{a15}) using the one-loop approximations for the matching functions $C_V$ and $\tilde s_{\rm DY}$ along with two-loop (three-loop) expressions for the (cusp) anomalous-dimension functions. At N$^3$LL order, one uses the two-loop approximations for the matching functions along with three-loop (four-loop) expressions for the (cusp) anomalous dimensions. Note that the two-loop virtual SUSY corrections $c_{V,{\rm SUSY}}^{(2)}$ in (\ref{a16}) and the four-loop cusp anomalous dimension are currently unknown, but their numerical impact on the N$^3$LL result is expected to be negligibly small. In our N$^3$LL results below, we include the known two-loop corrections to the hard and soft functions and the relevant three-loop anomalous dimensions.

Subleading terms can be added to our resummed expressions by matching them to fixed-order perturbation theory. To this end, we define
\be\label{a21}
   d\sigma^{\rm N^nLL+NLO} = d\sigma^{\rm N^nLL}|_{\mu_h,\mu_s,\mu_f} 
    + \Big( d\sigma^{\rm NLO}|_{\mu_f} - d\sigma^{\rm N^nLL}|_{\mu_h=\mu_s=\mu_f} \Big) 
    \bigg|_{{\cal O}(\alpha_s)} .
\ee
The first term on the right-hand side denotes the resummed prediction for the cross section, while the second one includes those terms that are subleading in the $z\to 1$ limit. They are obtained by subtracting the fixed-order expression for the leading singular terms, derived by setting all three scales equal in expression (\ref{a14}), from the complete fixed-order result. This difference is then expanded to first order in $\alpha_s$. Since the matching to fixed-order theory is somewhat cumbersome, we will sometimes restrict our analysis to the leading singular terms only. This will be mentioned explicitly below.

\subsection{Scale setting}
\label{subsec:scales}

\begin{figure}[t]
\begin{center}
\includegraphics[width=0.48\textwidth]{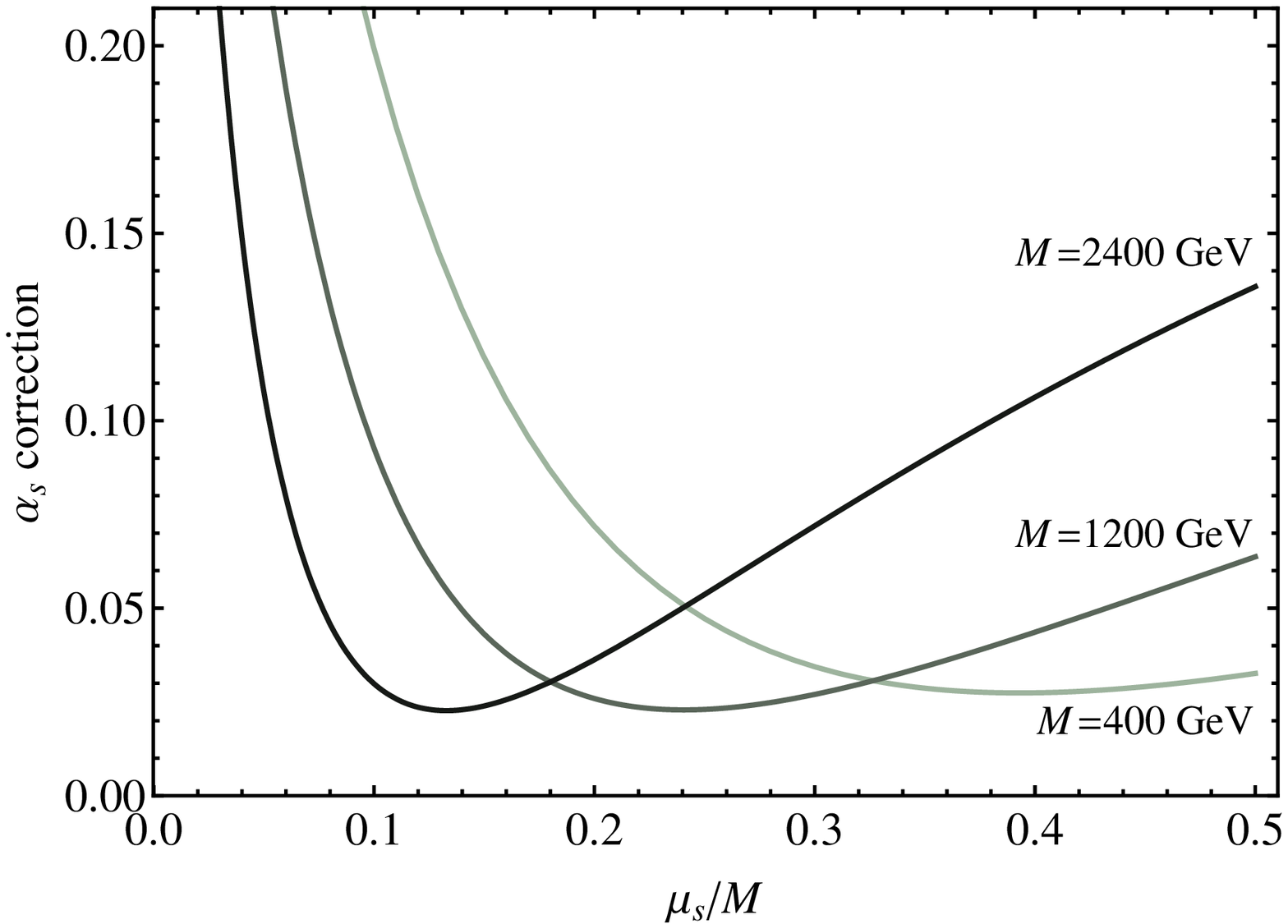}
\quad
\includegraphics[width=0.48\textwidth]{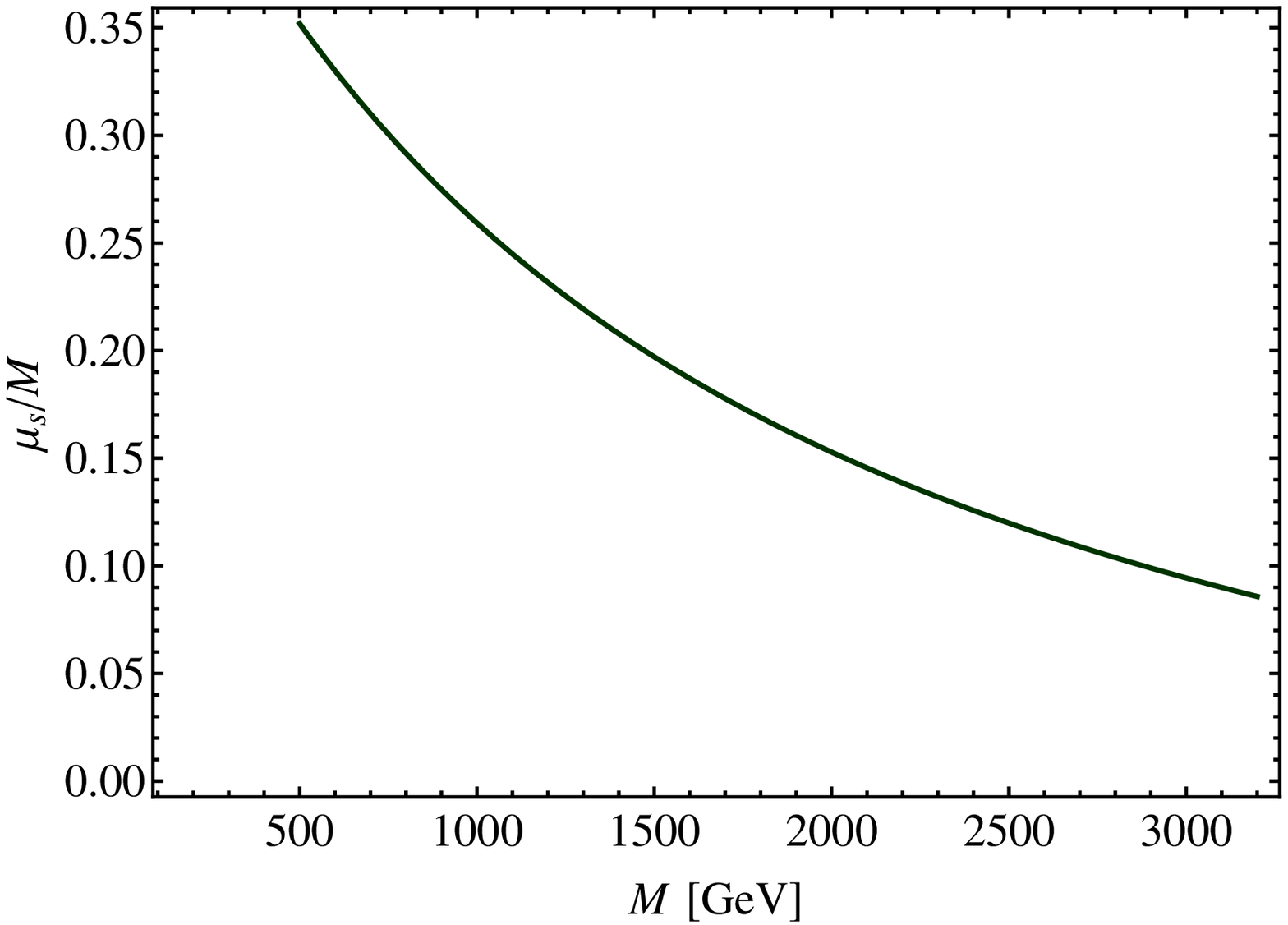}
\end{center}
\vspace{-4mm}
\caption{\label{fig4}
Relative contribution of the one-loop correction to the soft function to the cross section for slepton-pair production at the LHC ($\sqrt{s}=7$\,TeV), for different values of the pair invariant mass $M$ (left). For each value of $M$ we determine the soft scale by taking the point at which the correction is minimal (right).}
\end{figure}

An appropriate choice of the matching scales $\mu_h$ and $\mu_s$, which enter in the resummation formula (\ref{a14}), is crucial for the reduction of the remaining perturbative uncertainties in our calculation. In the spirit of effective field theory, the choice of these scales is driven by the requirement that the perturbative expansions of the matching coefficients $C_V$ and $\tilde s_{\rm DY}$ should be well behaved. Since the SUSY contributions to $C_V$ turn out to be very small (see below), these effects play no role in the scale-setting procedure, which therefore proceeds in analogy with the discussion for the Drell-Yan cross section presented in \cite{Becher:2007ty}. For the hard matching scale, we adopt the default choice $\mu_h^{\rm def}=M$. 

The soft scale $\mu_s$ is set dynamically by minimizing the effect of the one-loop corrections to the soft function $\tilde s_{\rm DY}$ under the convolution integral (\ref{a9}). This scale therefore depends on the value of $\tau=M^2/s$ and on the process under consideration. For the case of slepton-pair production at the LHC (with $\sqrt{s}=7$\,TeV), the result is shown in the left plot of Figure~\ref{fig4} for different choices of the invariant mass $M$. Our default value for the soft scale is chosen such that, for fixed $M$, the contribution of the one-loop correction to the soft function to the cross section is minimized. The value of $\mu_s/M$ for which this condition is satisfied is shown in the plot on the right. We observe that for sufficiently large values of $M$ the soft scale is indeed much smaller than the hard scale $\mu_h\sim M$, indicating the relevance of threshold resummation. The corresponding plots for the Tevatron and the LHC with $\sqrt{s}=14$\,TeV would look similar. For practical purposes, the values of $\mu_s$ as a function of $M$ and $s$ can be parametrized by means of the function
\be
   \mu_s^{\rm def} = \frac{M(1-\tau)}{\left(a+b\,\tau^{1/2}\right)^c} \,,
\ee
where $(a,b,c)=(1.1,3.6,1.9)$ for the Tevatron, $(1.5,4.8,1.7)$ for the LHC with $\sqrt{s}=7$\,TeV, and $(1.4,3.6,2.0)$ for the LHC with $\sqrt{s}=14$\,TeV. 

\begin{figure}[t]
\begin{center}
\includegraphics[width=0.48\textwidth]{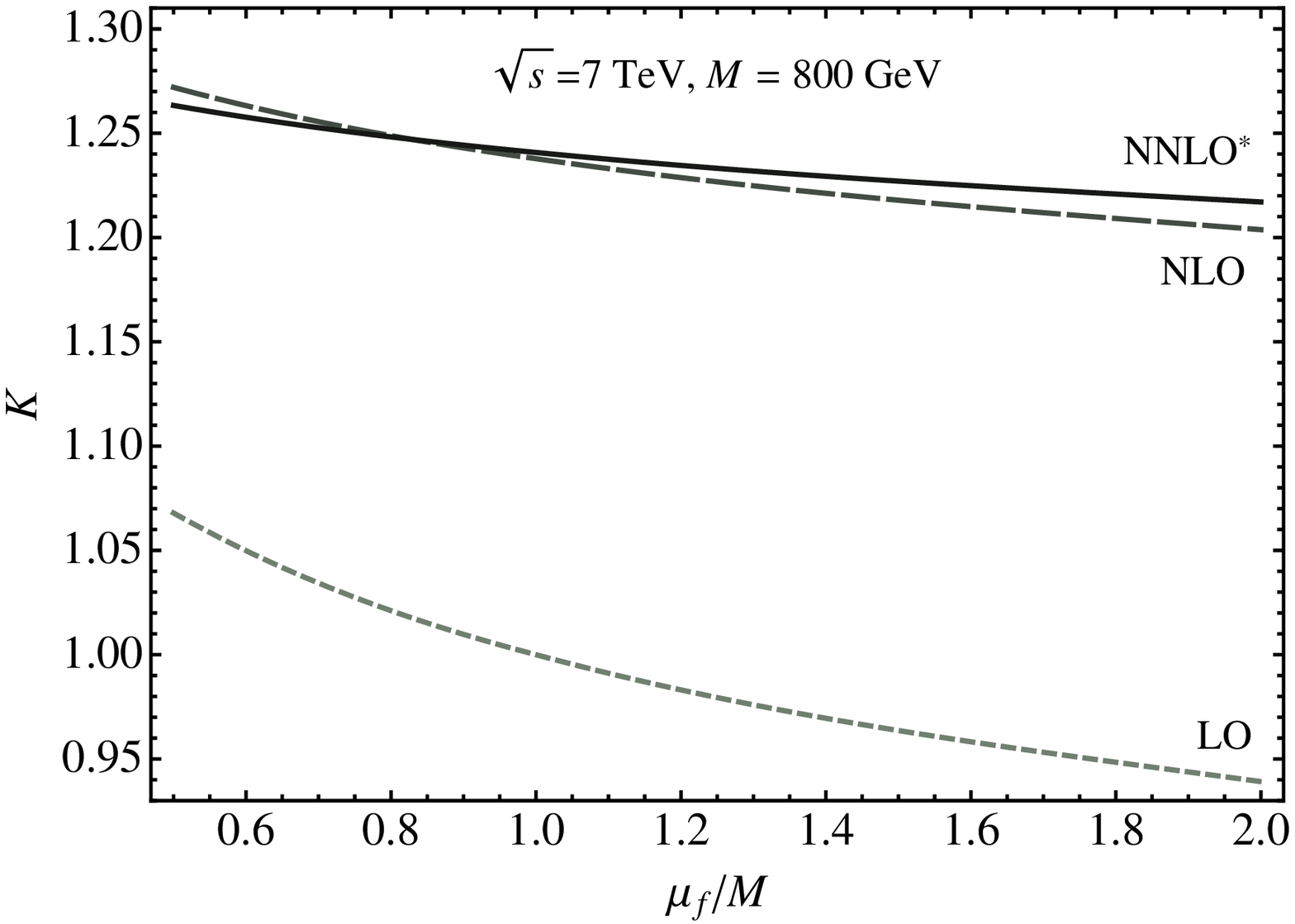}
\quad
\includegraphics[width=0.48\textwidth]{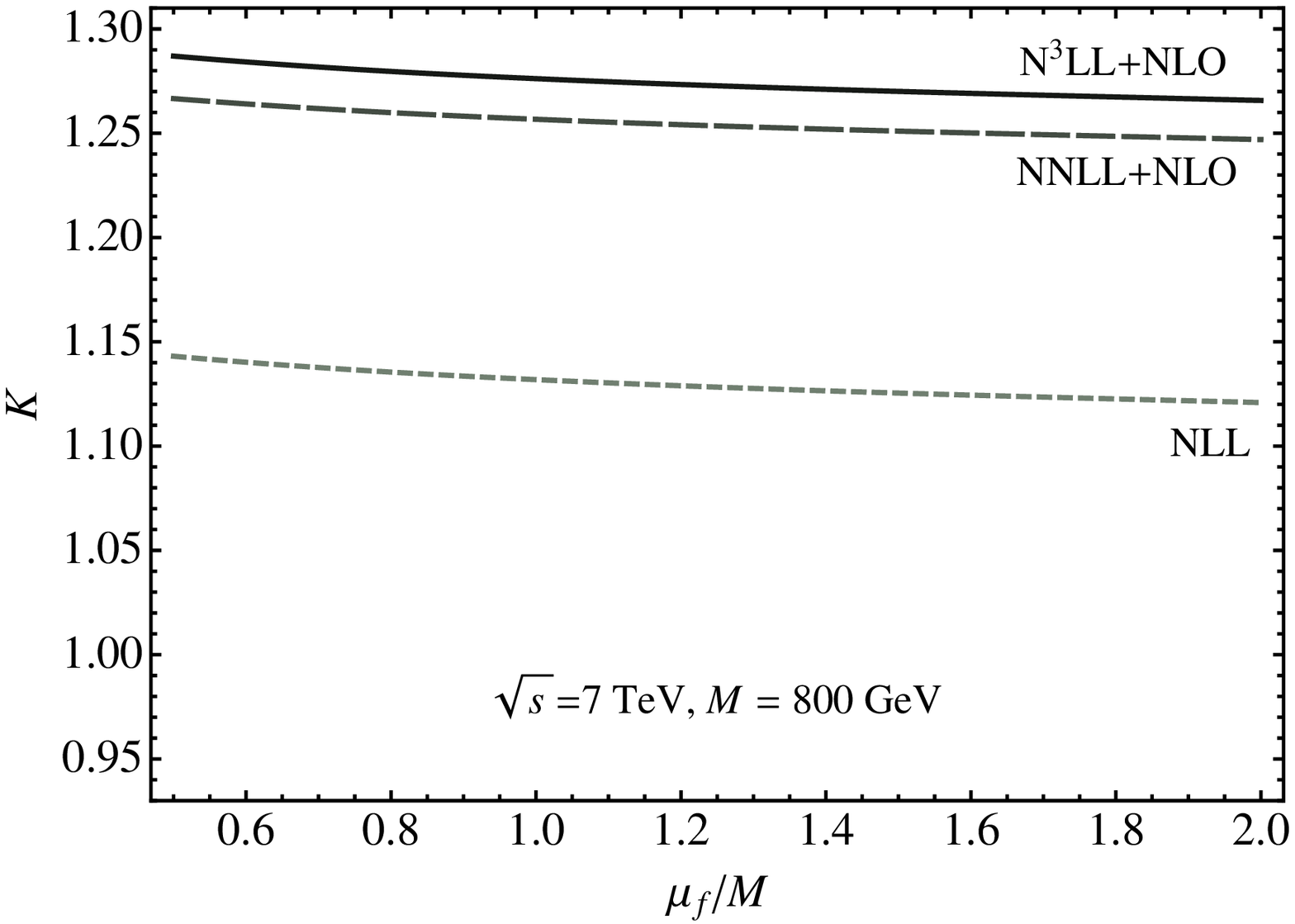}
\end{center}
\vspace{-5mm}
\caption{\label{fig6}
Factorization-scale dependence of the $K$-factor for slepton-pair production at the LHC in fixed-order perturbation theory (left) and after soft-gluon resummation (right). The NNLO$^*$ and N$^3$LL+NLO results contain only the leading singular two-loop corrections.}
\vspace{4mm}
\begin{center}
\includegraphics[width=0.47\textwidth]{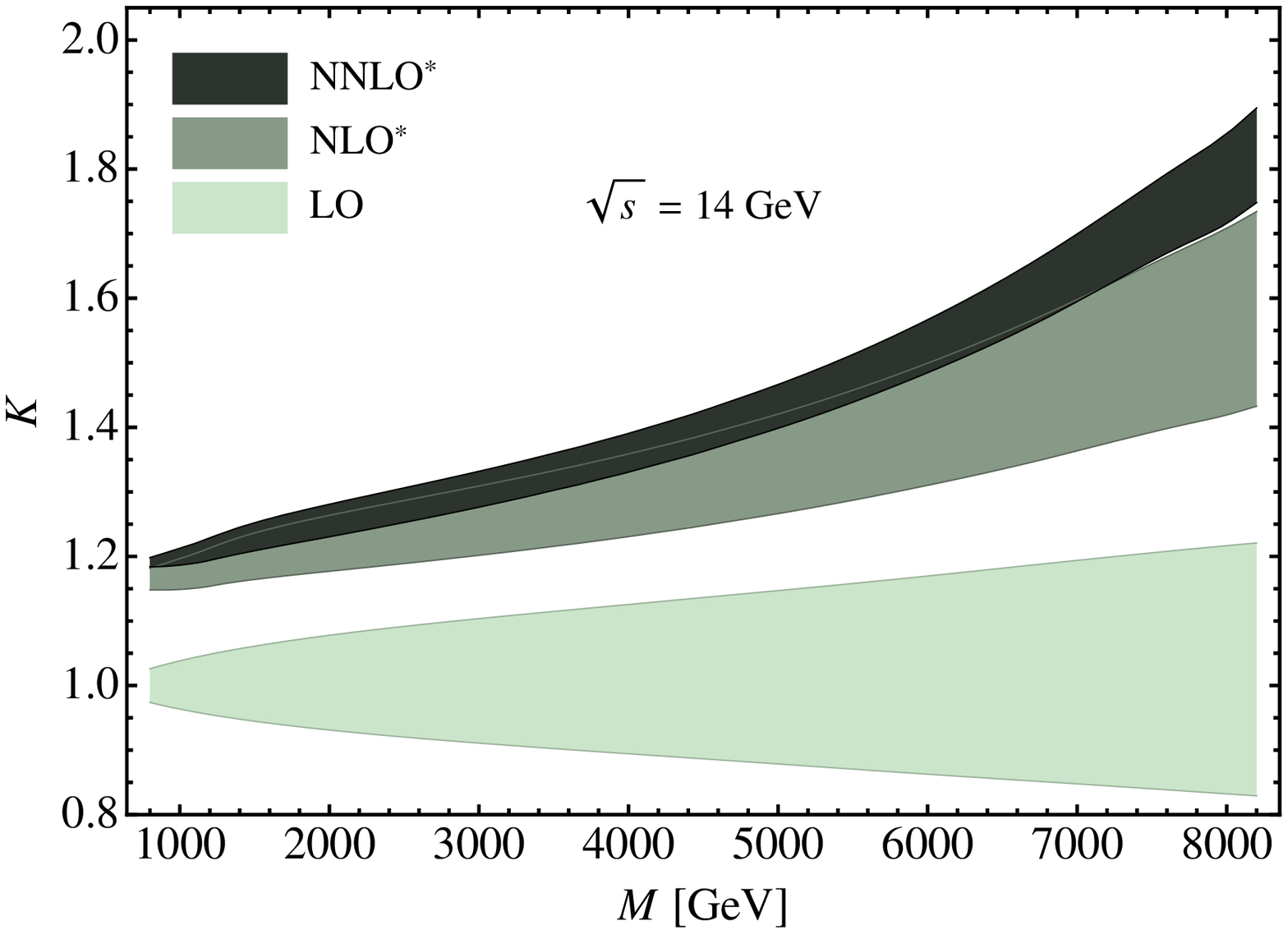}
\hspace{5mm}
\includegraphics[width=0.47\textwidth]{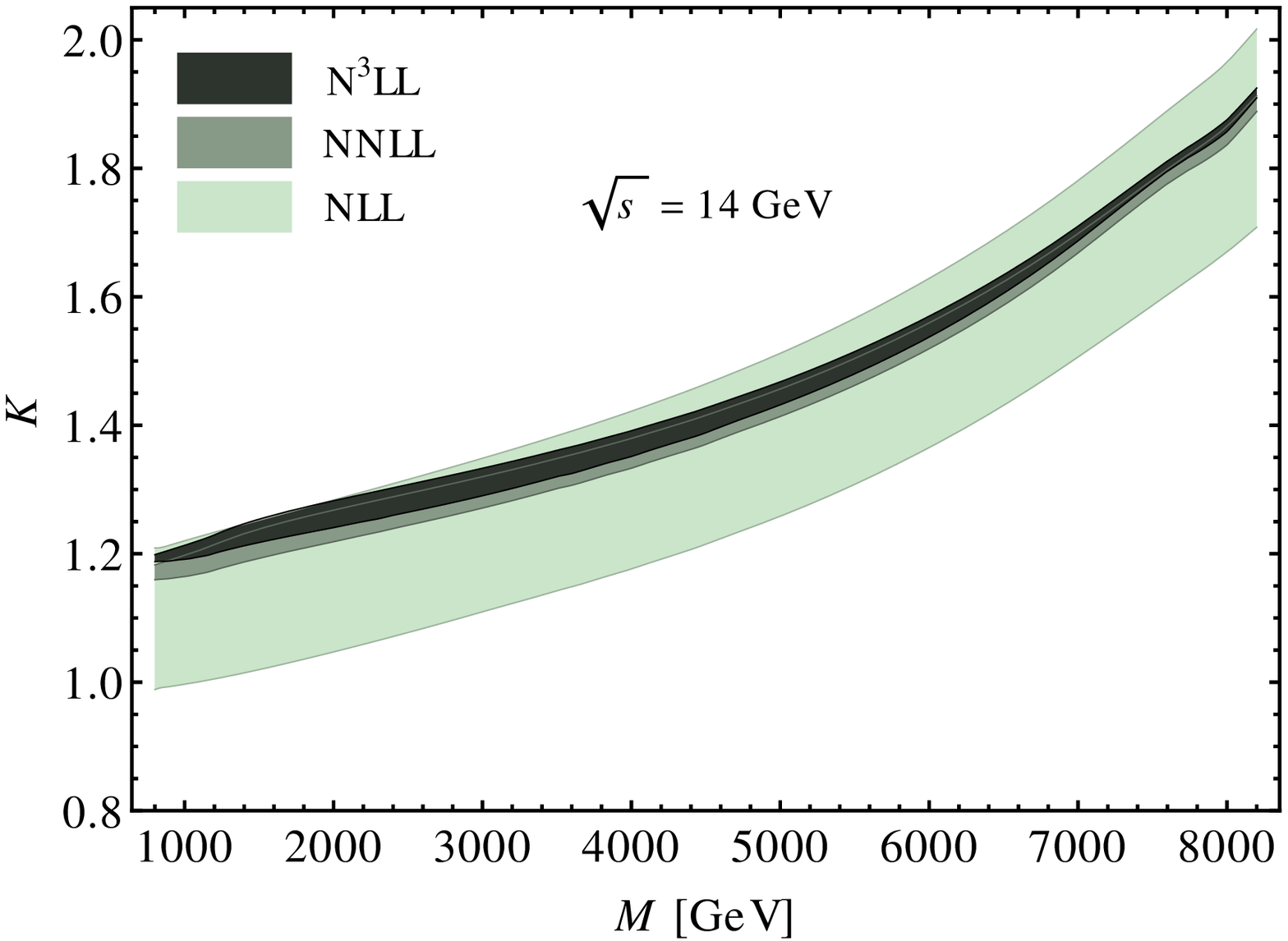}
\end{center}
\vspace{-5mm}
\caption{\label{fig7}
$K$-factor for slepton-pair production at the LHC, at different orders in fixed-order perturbation theory (left) and including the effects of soft-gluon resummation (right). Only the leading singular terms are included.}
\end{figure}

After the matching scales have been set, our results still exhibit a residual dependence on the factorization scale $\mu_f$, at which the PDFs are renormalized. As illustrated in Figure~\ref{fig6} for the case of the $K$-factor for slepton-pair production at the LHC, we find that after soft-gluon resummation this dependence is significantly weaker than in fixed-order perturbation theory, already at NLL order. For this analysis only the leading singular two-loop corrections are considered at NNLO, and this is denoted by a star (NNLO$^*$). It follows that there is very little sensitivity to the choice of the factorization scale. Below we take $\mu_f^{\rm def}=M$ as our default value.

In Figure~\ref{fig7}, we study the $K$-factor for slepton-pair production at the LHC, showing results obtained at different orders of perturbation theory. Only the leading singular terms are included in all cases. The widths of the various bands reflect the scale uncertainties inherent in the calculations. For the fixed-order results, they are obtained by setting the renormalization scale $\mu_r$ equal to the factorization scale $\mu_f$, and varying the common scale between 0.5 and 2 times its default value. We performed an independent variation of the two scales as well, but the corresponding uncertainty obtained by taking an envelope of the maximum deviation from the default value does not differ appreciably from the result in Figure~\ref{fig7}. This is because at leading order the cross section depends on $\mu_f$ only, while the dependence on $\mu_r$ starts at NLO and is therefore small. For the resummed results, the error bands take into account the uncertainties associated with the scales $\mu_h$, $\mu_s$, and $\mu_f$. They correspond to the envelope of the predictions obtained by varying all three scales simultaneously between 0.5 and 2 times their default values. We verified that the bands obtained in this way do not differ in a significant way from those obtained by varying the three scales individually, with the other two scales held fixed, and adding then the three uncertainties in quadrature. It is evident that the convergence of the perturbative expansion and the remaining scale uncertainties are greatly improved by means of soft-gluon resummation. After resummation the three bands nicely overlap, and the scale uncertainties are reduced significantly with each order. On the contrary, the fixed-order results exhibit a slower convergence and larger scale uncertainties. 

\subsection{Impact of SUSY matching corrections}
\label{subsec:susy}

The existence of SUSY particles would affect our analysis in two ways. First, if sleptons exist and are kinematically accessible at the Tevatron or LHC, these particles can be pair produced, and via a measurement of their cross section and invariant-mass distribution one can address questions about the slepton masses and couplings. Secondly, strongly-interacting SUSY particles (squarks and gluinos) can affect the hard-scattering kernels for both lepton-pair and slepton-pair production at ${\cal O}(\alpha_s)$, via one-loop SUSY QCD corrections to the hard matching coefficient $C_V$ in (\ref{a15}). This second effect is obviously more subtle than the first one. We will now address if and to what extent it will be possible to probe for virtual effects of SUSY particles in high-precision measurements of the standard Drell-Yan cross section. 

\begin{figure}[t]
\begin{center}
\hspace{2mm}
\includegraphics[width=0.45\textwidth]{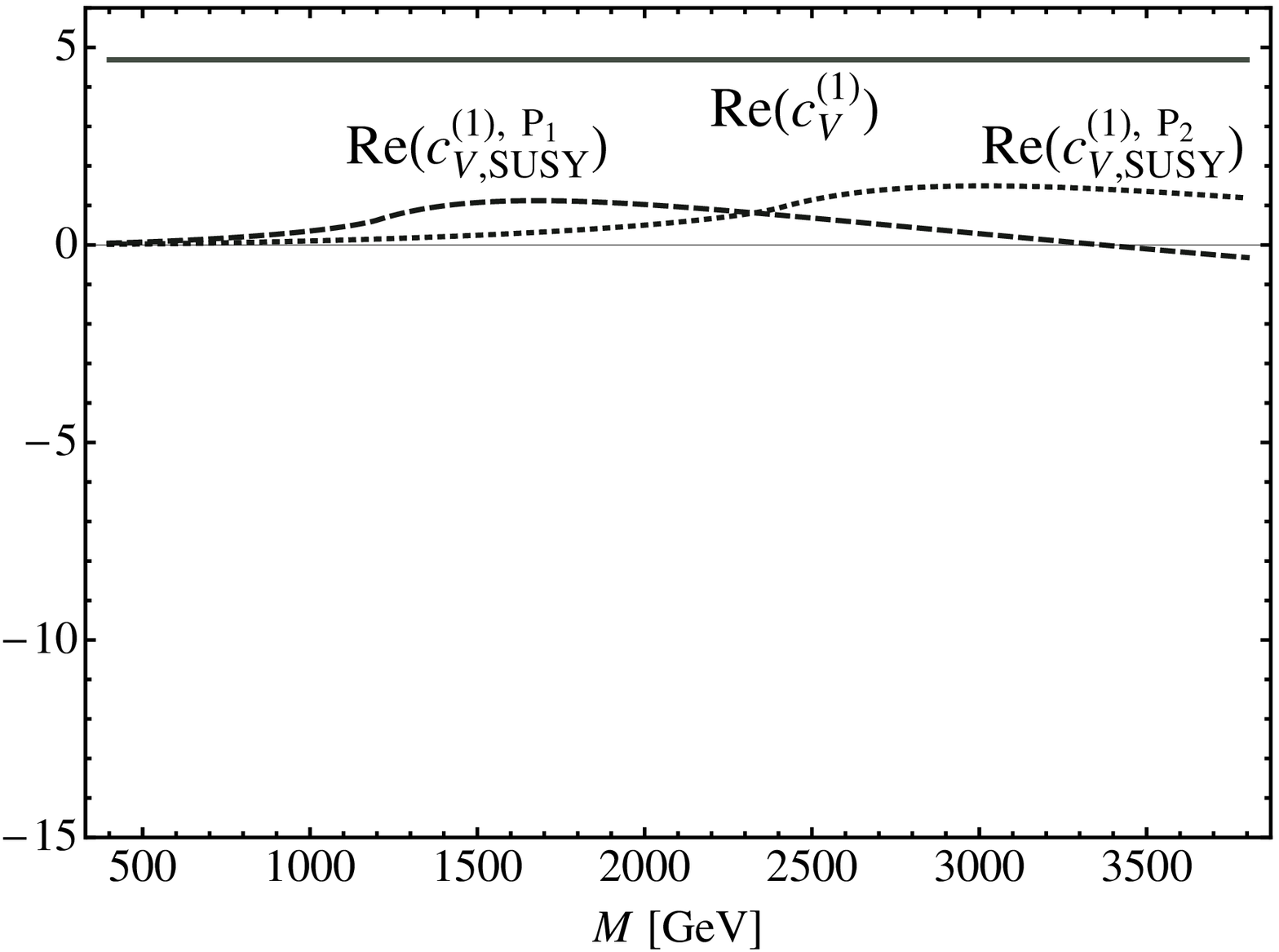}
\hspace{5mm}
\includegraphics[width=0.45\textwidth]{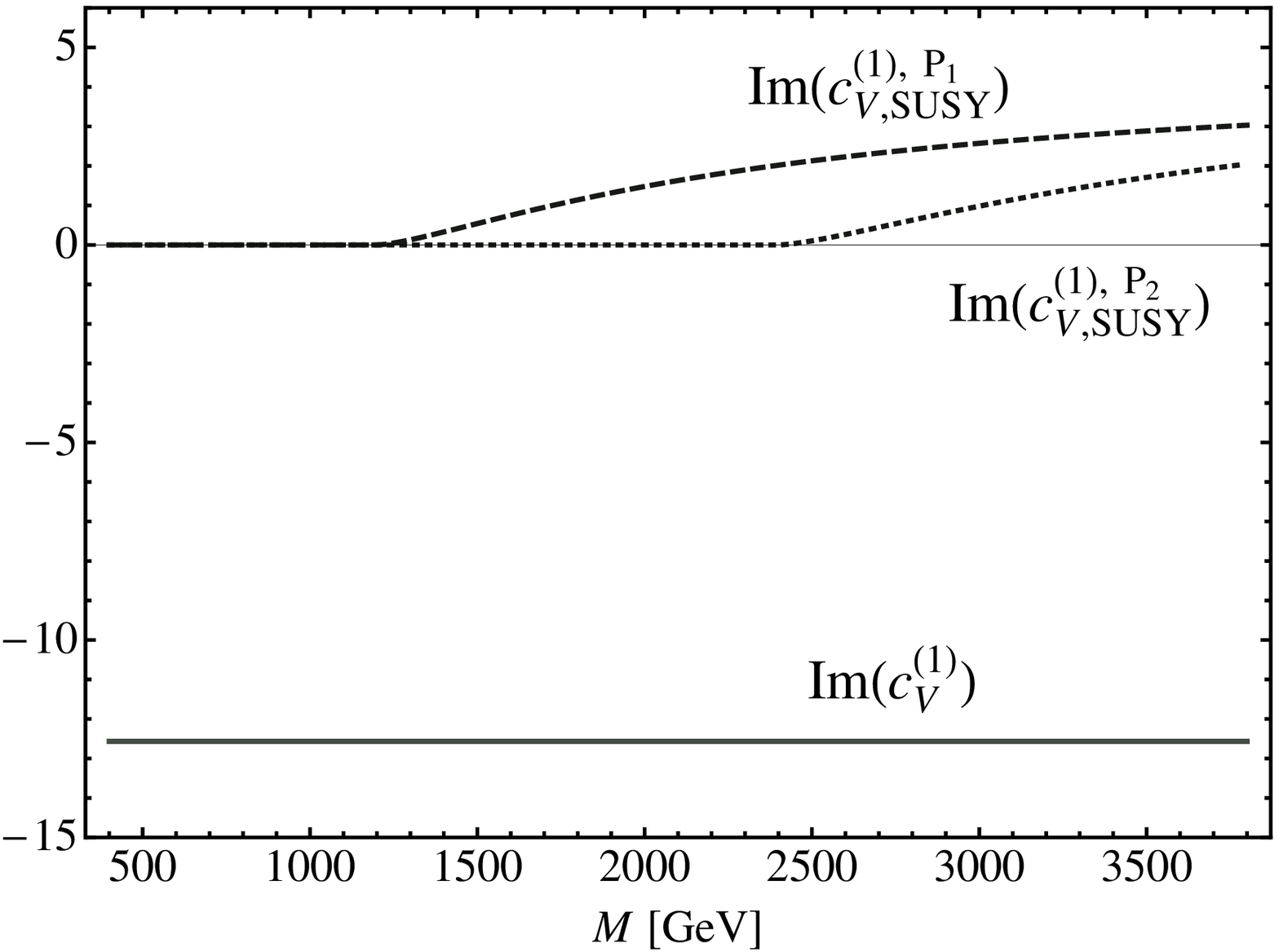}
\end{center}
\vspace{-5mm}
\caption{\label{fig:c1}
Comparison of one-loop contributions to the hard matching coefficient $C_V$ arising in the SM (solid lines) and in its SUSY extensions (dashed lines).}
\vspace{4mm}
\begin{center}
\includegraphics[width=0.47\textwidth]{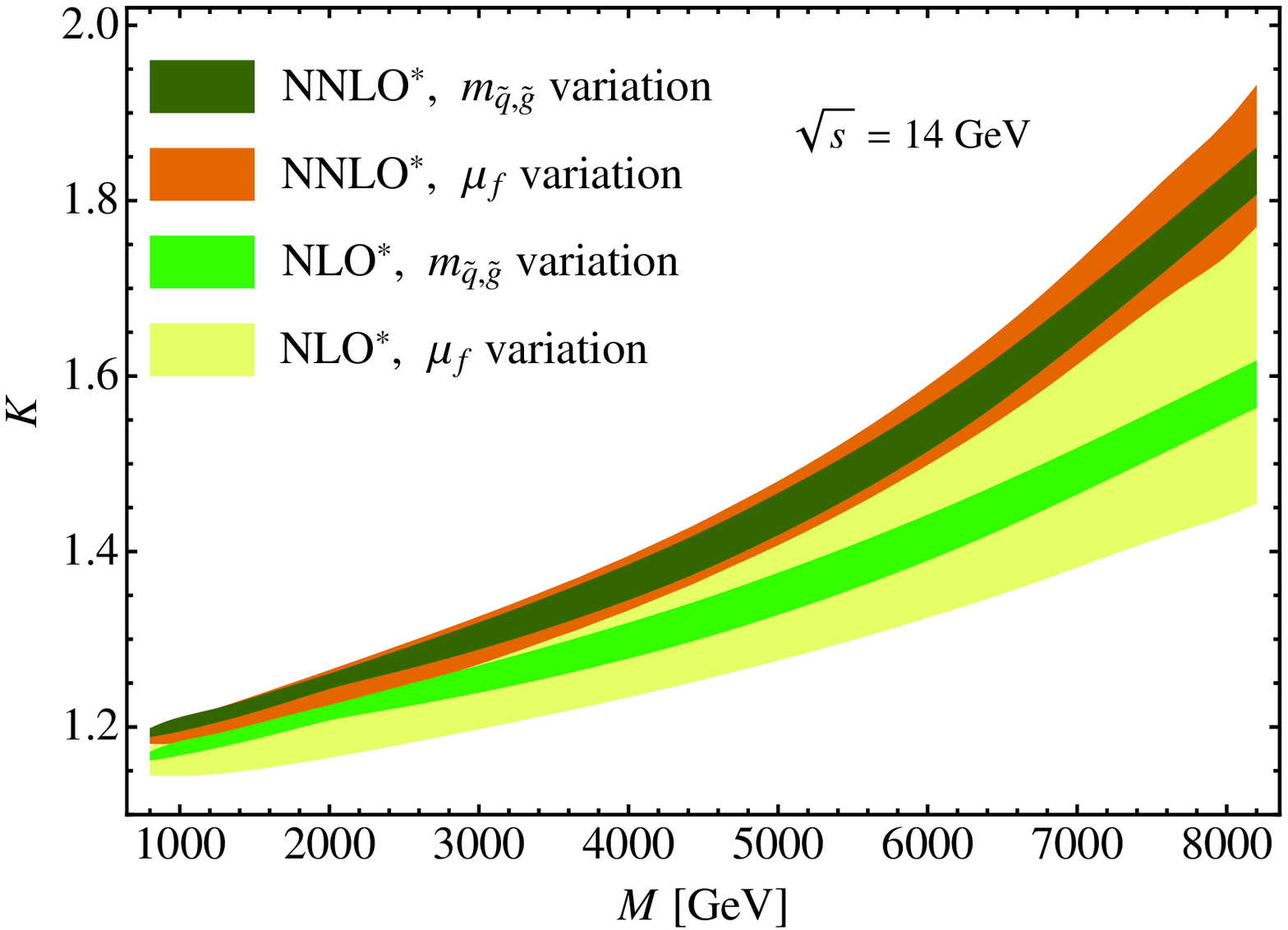}
\quad
\includegraphics[width=0.47\textwidth]{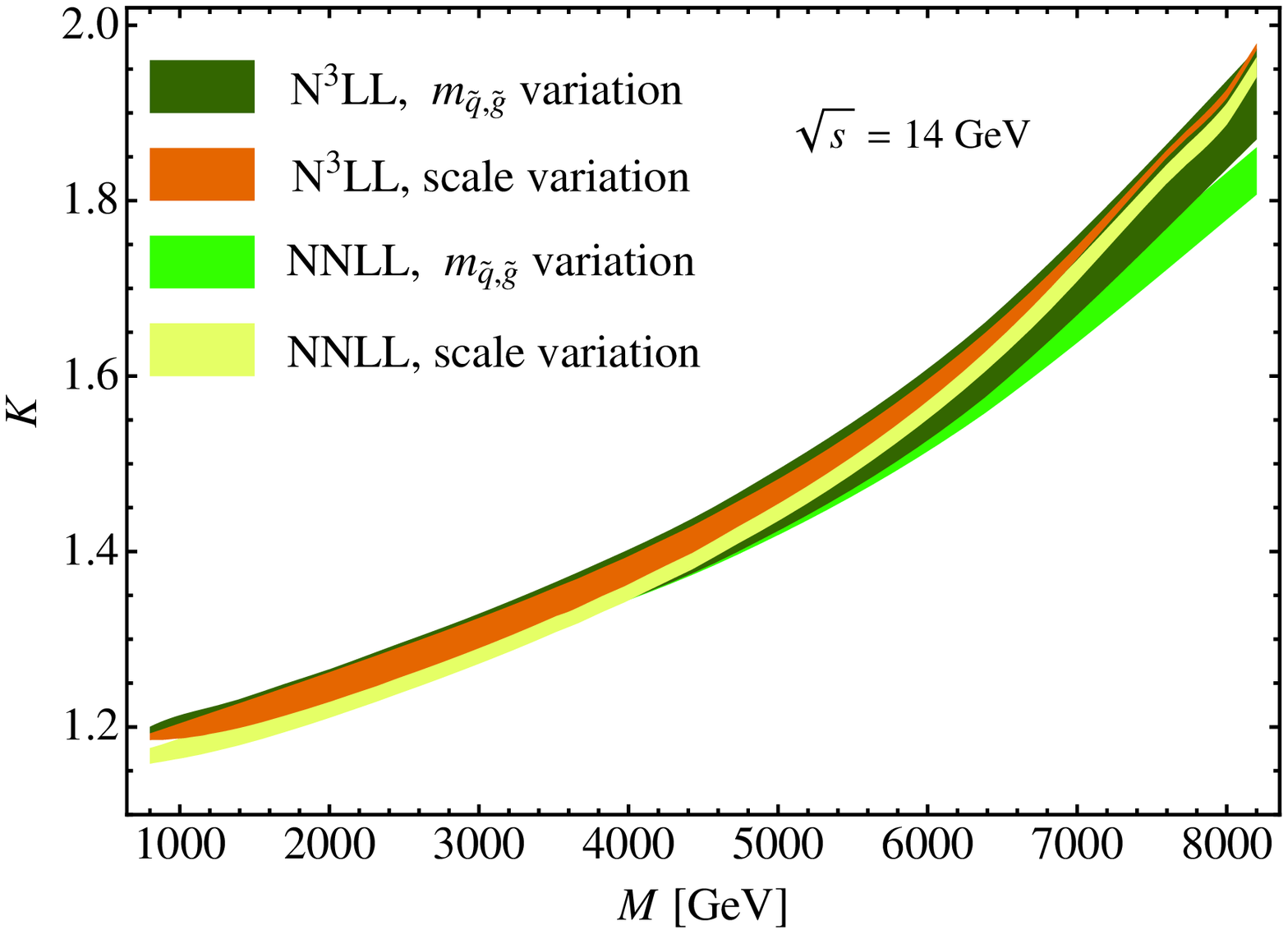}
\end{center}
\vspace{-5mm}
\caption{\label{fig3}
Comparison of the scale uncertainty of the Drell-Yan $K$-factor for the LHC (light bands) with the maximum possible deviation due to loop effects of SUSY particles (dark bands). The left plot refers to fixed-order perturbation theory, while the right one includes the effects of soft-gluon resummation. Only the leading singular terms are included in the calculation.}
\end{figure}

In Figure~\ref{fig:c1}, we compare the one-loop SUSY QCD contributions to the Wilson coefficient $C_V$ with the corresponding QCD contributions arising in the SM, for two representative choices of the squark and gluino masses. The first, $m_{\tilde q}=600$\,GeV and $m_{\tilde g}=750$\,GeV (parameter point $P_1$), represents an average value for the squark and gluino masses close to the point SPS1a' \cite{AguilarSaavedra:2005pw}. The second, given by $m_{\tilde q}=1200$\,GeV and $m_{\tilde g}=500$\,GeV (parameter point $P_2$), represents an alternative scenario with a lighter gluino and heavier squarks, inspired by scenarios like SPS2 and SPS3 \cite{Allanach:2002nj}. Although close to current limits on sparticle masses \cite{Chatrchyan:2011zy,Aad:2011ib,Aad:2011cw,Allanach:2011qr,Buchmueller:2011sw}, these points are not yet ruled out in the context of a general SUSY extension of the SM. We observe that for all parameter choices the SM loop corrections are at least a factor of about 3 larger in magnitude than the SUSY corrections. The real part of the SUSY contributions reaches a maximum close to the squark threshold, for $M\gtrsim 2 m_{\tilde q}$, where the virtual contributions acquire an imaginary part. For values of $M$ below threshold the SUSY corrections are yet much smaller. As a result, we find that the virtual SUSY effects are smaller than the residual scale uncertainties in the calculation of the cross sections. This is shown in Figure~\ref{fig3}, where we compare the factorization-scale uncertainty (both at fixed order and after resummation) of the SM-only Drell-Yan $K$-factor with the maximum deviation arising if we include the virtual SUSY contributions and scan the squark and gluino masses independently over the range between 400 and 2200\,GeV. For each value of $M$ we choose the points in the ($m_{\tilde q},m_{\tilde g}$) plane which yield the largest up- and downward deviations. In the figure we consider Drell-Yan production at the LHC with $\sqrt{s}=14$\,TeV. At lower energy or for the Tevatron the impact of virtual SUSY effects is even smaller. We observe that in fixed-order perturbation theory the maximum possible deviation due to virtual effects of SUSY particles is always much smaller than the residual scale dependence. After soft-gluon resummation the size of the effects is comparable; however, in view of the fact that additional theoretical uncertainties arise from the variation of the PDFs (not shown in the figure), we conclude that also in this case the combined theoretical error is larger than the maximum possible SUSY effect.

\begin{figure}[t]
\begin{center}
\includegraphics[width=0.98\textwidth]{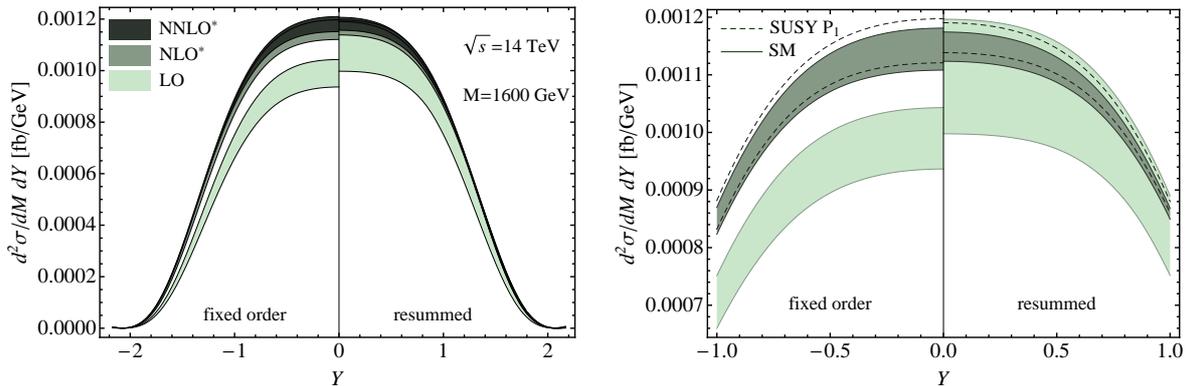}
\end{center}
\vspace{-6mm}
\caption{\label{fig8}
Drell-Yan rapidity distribution at fixed $M=1600$\,GeV at different orders in perturbation theory. In both plots, the left half refers to fixed-order perturbation theory, while the right half includes the effects of soft-gluon resummation. The plot on the left refers to the SM, while the one on the right shows the effects of virtual squarks and gluinos by the dashed bands.}
\end{figure}

On the one hand, these findings justify an approximation at N$^3$LL order where in the two-loop corrections to the hard matching coefficient $C_V$ in (\ref{a15}) one neglects the SUSY contribution $c_{V,\rm SUSY}^{(2)}$ compared with the corresponding SM contribution $c_V^{(2)}$. This approximation will be adopted in our N$^3$LL predictions below. On the other hand, it appears unlikely that it will be possible to probe for SUSY effects via virtual corrections to the Drell-Yan cross section. To illustrate this latter point, we show in Figure~\ref{fig8} the result for the Drell-Yan rapidity distribution in our SUSY model $P_1$ (with $m_{\tilde q}=600$\,GeV and $m_{\tilde g}=750$\,GeV) at a value of the pair invariant mass for which the SUSY effects are close to maximal (cf.\ Figure~\ref{fig:c1}). We consider the LHC with $\sqrt{s}=14$\,TeV and restrict our analysis to the leading singular terms only. In the left plot we compare the results obtained at different orders in fixed-order perturbation theory with the corresponding results obtained after soft-gluon resummation. These distributions refer to the SM without SUSY effects. We observe that resummation improves the convergence of the perturbative expansion and leads to somewhat smaller scale variations at higher orders. The right plot zooms in on the central region of the rapidity distribution and shows once again the results obtained at leading and next-to-leading order with and without resummation. The dashed lines indicate the shift of the NLO$^*$ and NNLL bands due to the presence of SUSY particles. In both cases (with and without resummation) the shift amounts to a small enhancement of the cross section, which however is only a fraction of the residual scale uncertainty indicated by the widths of the bands. Note also that the additional PDF uncertainty is not included in the plots. 

\subsection{Invariant-mass distribution for slepton-pair production}

A clear signal of SUSY would come from the direct production and detection of slepton pairs. In this case, the very weak dependence on the squark and gluino masses can be seen as an advantage, since it would make it possible to use a measurement of the slepton-pair production cross section to extract the mass of the slepton produced, or alternatively to set a limit on the slepton mass from an upper limit on the production cross section. The sensitivity of the cross section (\ref{a1}) to the slepton mass $m_{\tilde l}$ arises from the prefactor $\beta_{\tilde l}^3$ in (\ref{a2}), and from the fact that the peak of the invariant mass distribution scales with $m_{\tilde l}$. After the determination of the matching and factorization scales in Section~\ref{subsec:scales}, we are now ready to analyze the impact of soft-gluon resummation on the invariant-mass distribution for slepton-pair production. For concreteness, we will consider the production of a pair of scalar leptons $\tilde l_L$, as this case has the largest cross section. For the SUSY masses we take $m_{\tilde l_L}=180$\,GeV, $m_{\tilde q}=600$\,GeV, and $m_{\tilde q}=750$\,GeV. 

\begin{figure}[p]
\begin{center}
\includegraphics[width=0.48\textwidth]{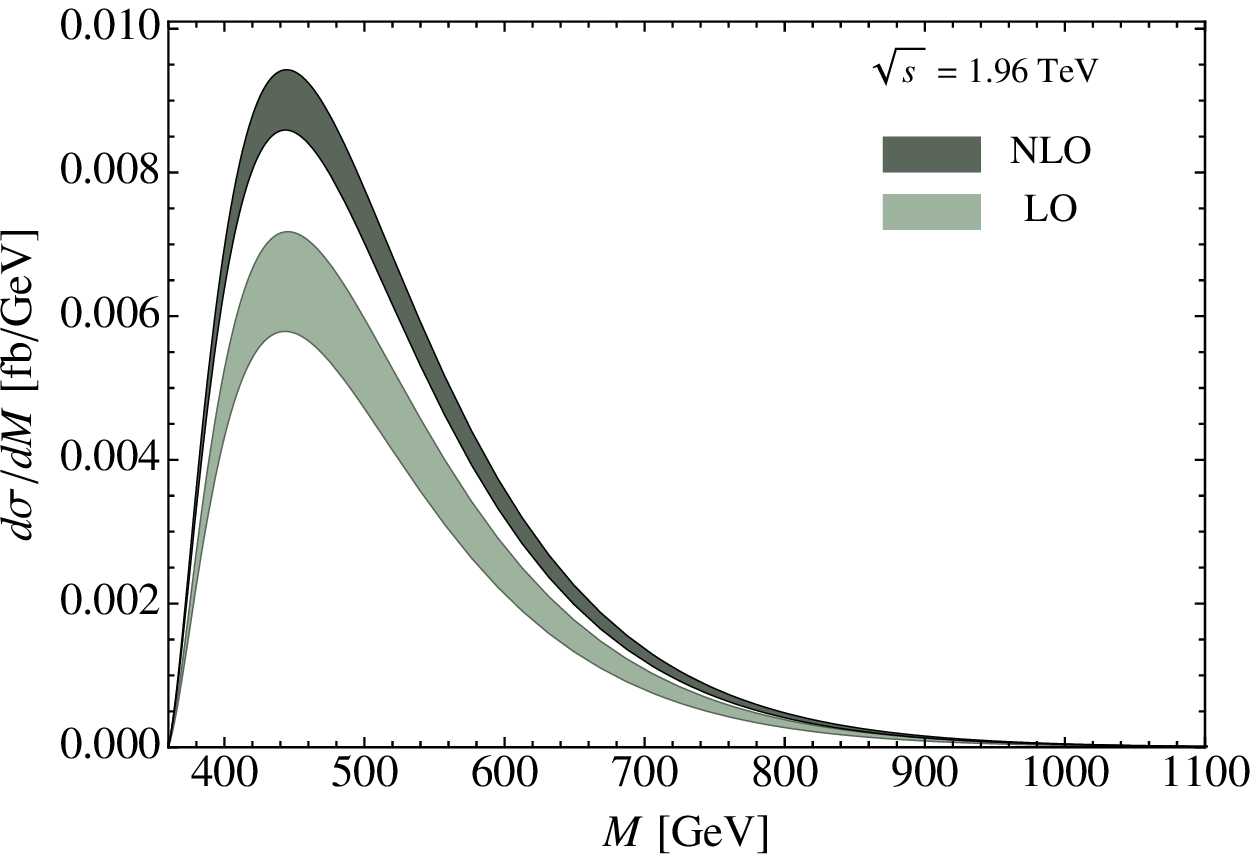}
\quad
\includegraphics[width=0.48\textwidth]{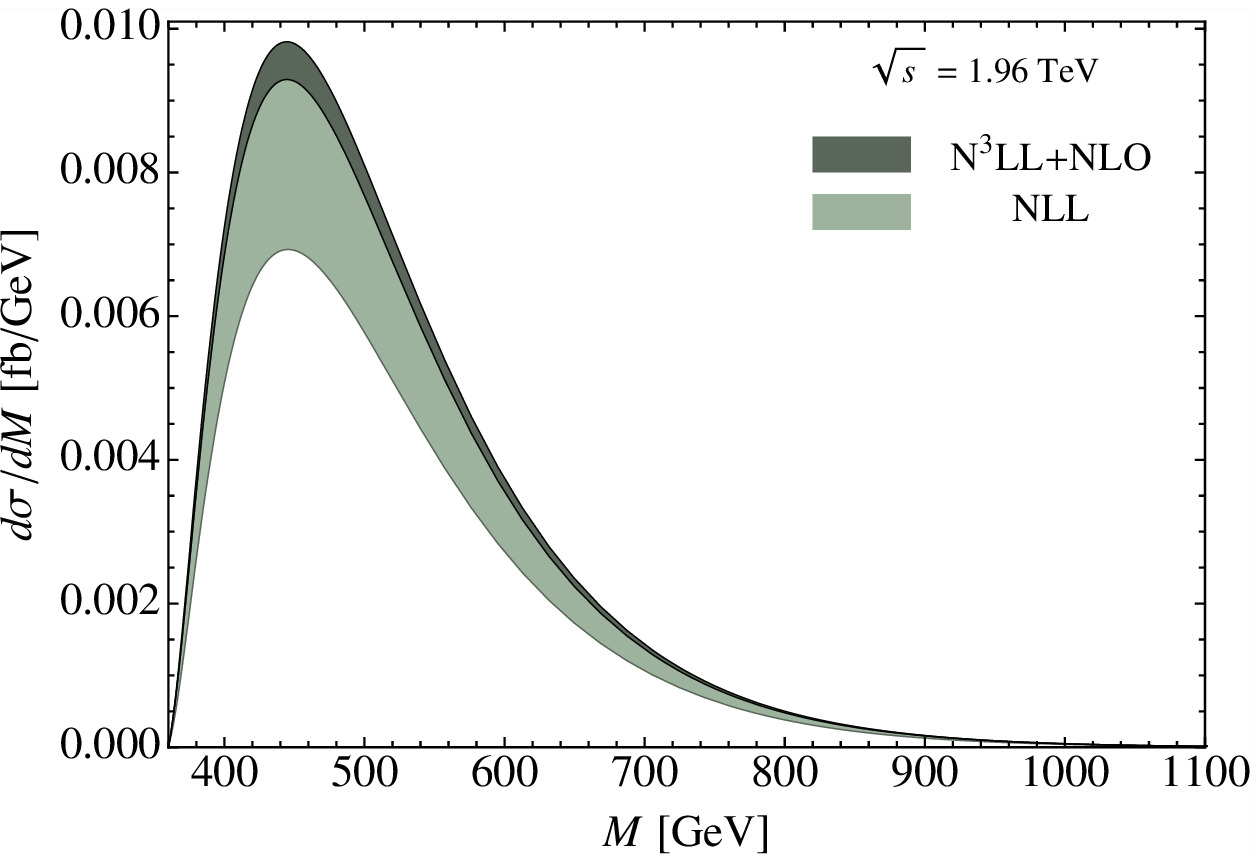}
\includegraphics[width=0.48\textwidth]{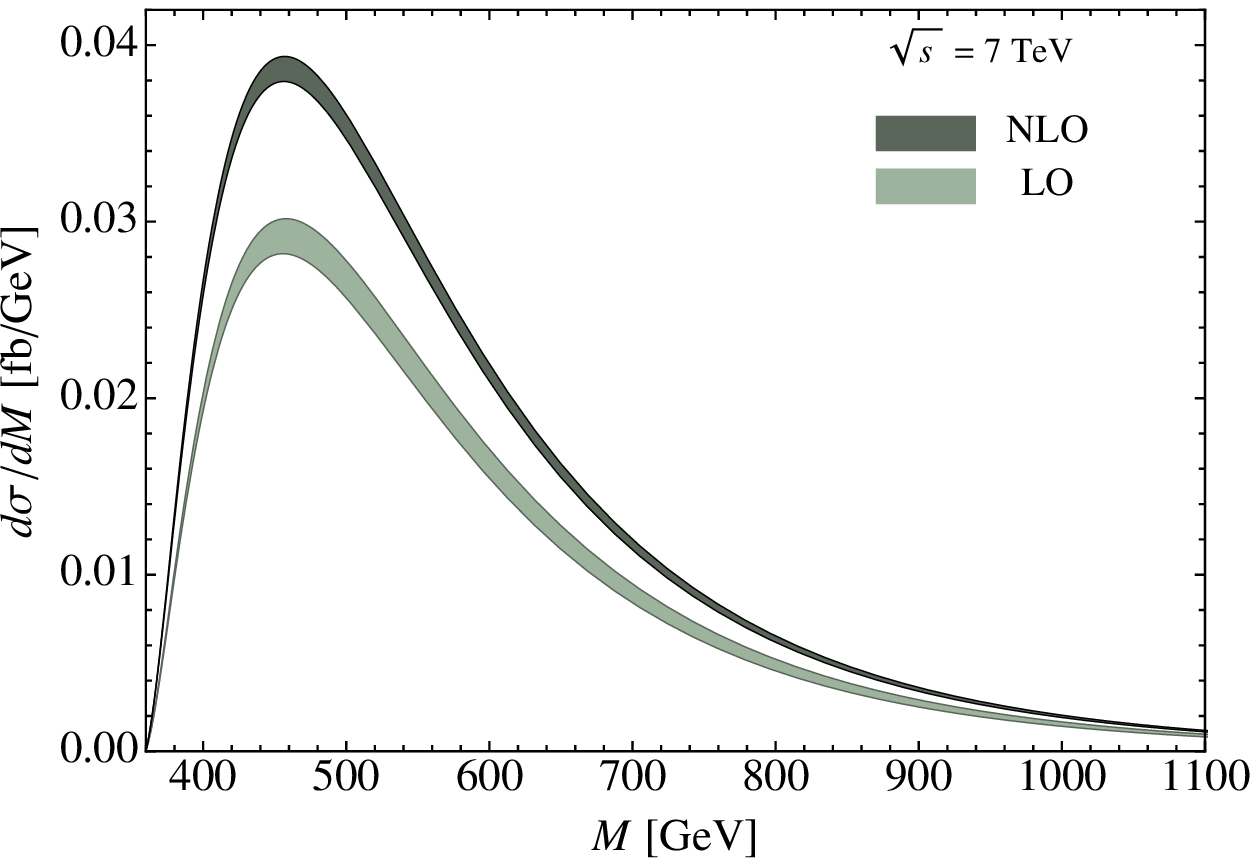}
\quad
\includegraphics[width=0.48\textwidth]{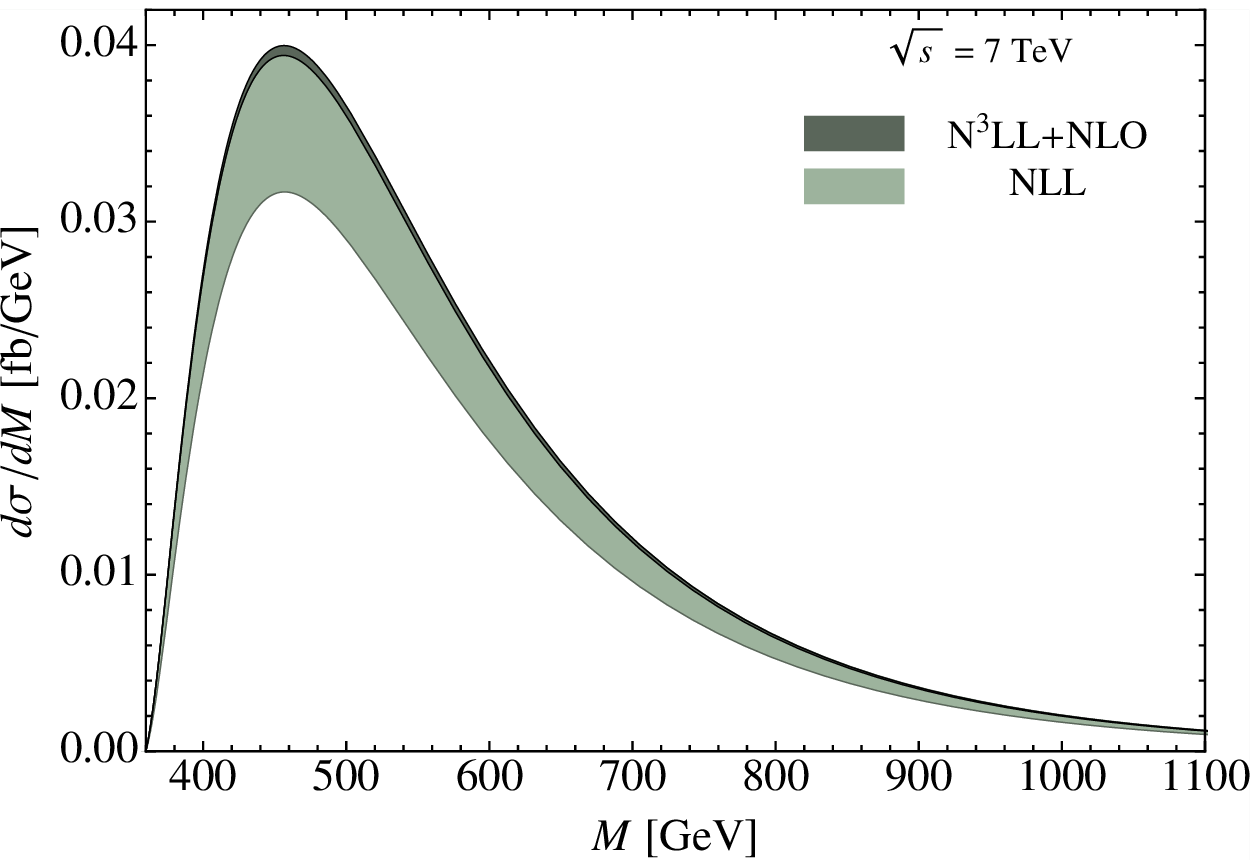}
\includegraphics[width=0.48\textwidth]{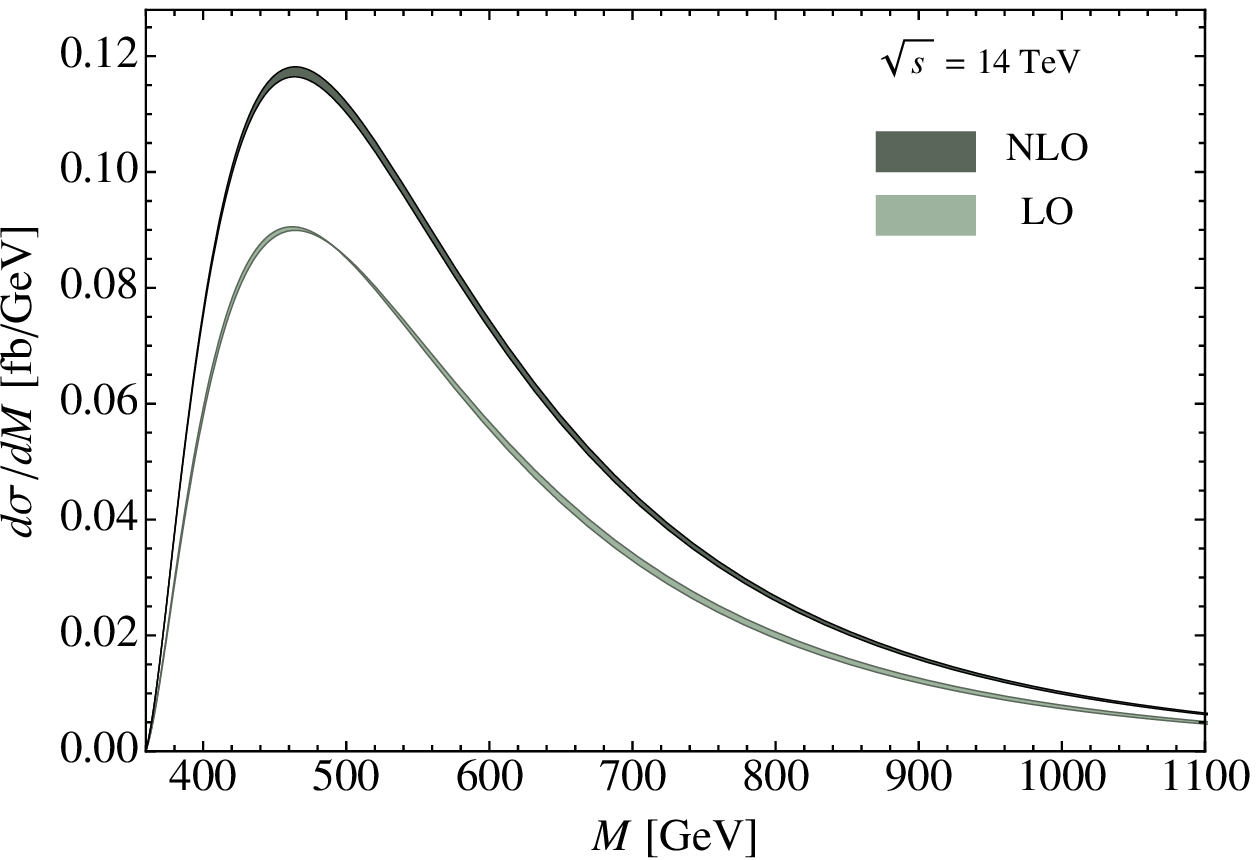}
\quad
\includegraphics[width=0.48\textwidth]{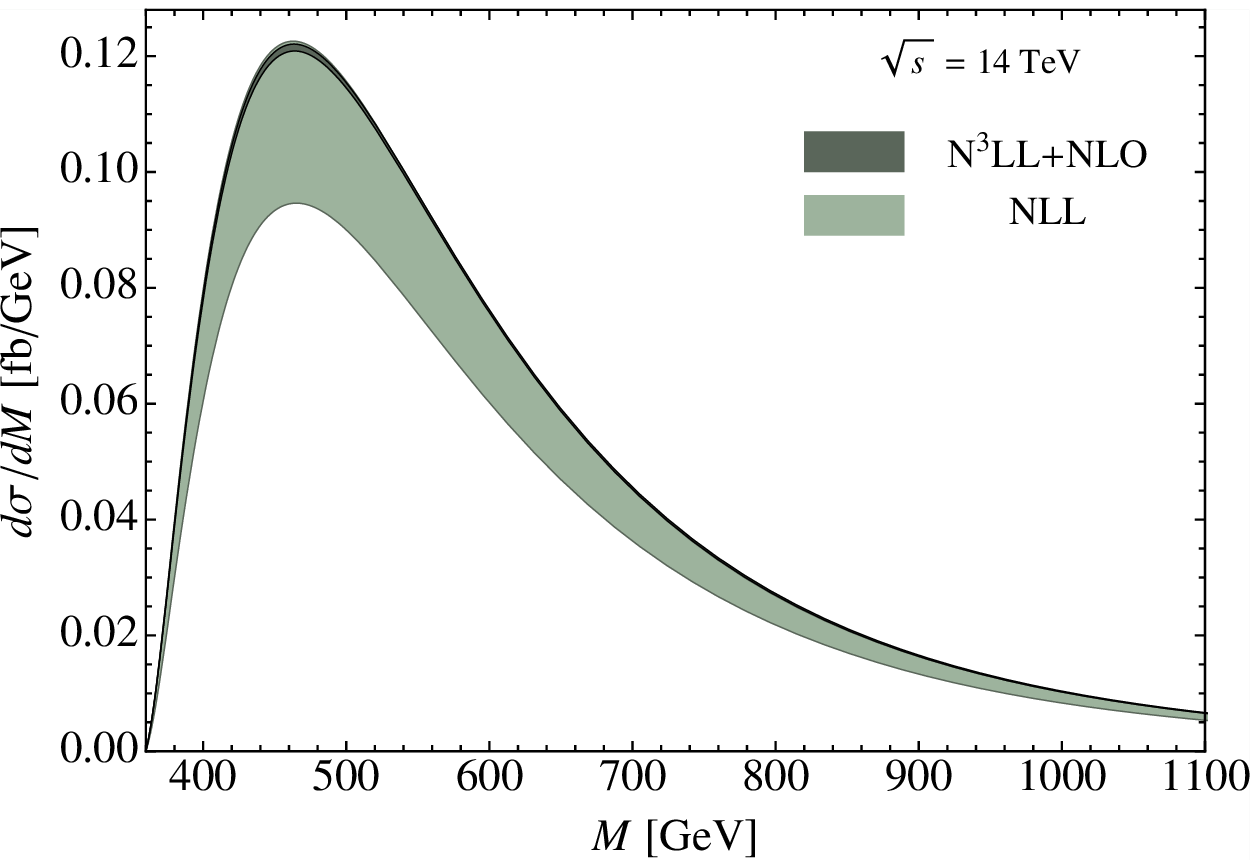}
\end{center}
\vspace{-4mm}
\caption{\label{fig9}
Invariant-mass distributions for slepton-pair production at the Tevatron and LHC. The plots on the left show fixed-order results at LO and NLO, while those on the right include the effects of soft-gluon resummation at NLL and N$^3$LL+NLO. The bands indicate the uncertainty associated with scale variations.}
\end{figure}

Our results for the slepton invariant-mass distributions are shown in Figure~\ref{fig9} for the Tevatron (top), the LHC with $\sqrt{s}=7$\,TeV (center), and the LHC with $\sqrt{s}=14$\,TeV (bottom). Contrary to our previous treatment, from now on we consider the PDFs and $\as$ at the order which is appropriate for the expansion of the corresponding hard-scattering kernels. Specifically, we use LO and NLO PDFs for the LO and NLO fixed-order results, and NLO and NNLO PDFs for the resummed results at NLL, NNLL+NLO, and N$^3$LL+NLO, since in this case the resummed terms include the bulk of the perturbative corrections appearing at one order higher in $\as$. Numerically, the NNLL+NLO (not shown in the figure) and N$^3$LL+NLO results turn out to be very close to each other, but the scale dependence of the latter ones is further reduced. We thus consider the N$^3$LL+NLO approximation as our best prediction. The main effect of soft-gluon resummation is to increase the cross sections slightly and to improve the convergence of the expansion. The resummation effects become more relevant for larger invariant masses. For example, at the Tevatron the increase from NLO to N$^3$LL+NLO is 7\% at 500\,GeV and 13\% at 1000\,GeV. The corresponding increases at the LHC are around 2\% for both $\sqrt{s}=7$\,TeV and 14\,TeV, indicating that at the LHC resummation effects are less important.

\subsection{Total cross section}

We obtain the total cross sections by integrating the invariant-mass distributions over $M$. In Table~\ref{tab2} we present results corresponding to the various approximations discussed in the previous section. In this section we provide predictions for both the NNLL+NLO and N$^3$LL+NLO total cross sections, so that the effect of including higher-order logarithmic terms can be seen. In the table the first error refers to the total scale variation, i.e.\ the variation of $\mu_f$ for the fixed-order cross sections and the maximum deviation from the default value obtained by varying $\mu_f$, $\mu_h$, and $\mu_s$ simultaneously  for the resummed and matched results. The second error takes into account the uncertainty of the PDFs at the 90\% confidence level, which is estimated by evaluating the cross sections with the 40 sets of PDFs provided by MSTW2008.

\begin{table}[t]
\centering
\begin{tabular}{|l|c|c|}
\hline
 & Tevatron (SUSY point $P_1$) & LHC (7\,TeV, SUSY point $P_1$) \\
\hline
$\sigma_{\rm LO}$ & $1.31_{-0.14}^{+0.17}\,_{-0.06}^{+0.08}$
 & $8.01_{-0.36}^{+0.39}\,_{-0.34}^{+0.31}$ \\
$\sigma_{\rm NLL}$ & $1.65_{-0.20}^{+0.27}\,_{-0.08}^{+0.12}$
 & $9.59_{-0.92}^{+1.25}\,_{-0.37}^{+0.41}$ \\
$\sigma_{\rm NLO}$ & $1.83_{-0.10}^{+0.09}\,_{-0.10}^{+0.14}$
 & $10.56_{-0.22}^{+0.24}\,_{-0.43}^{+0.48}$ \\
$\sigma_{\rm NNLL+NLO}$ & $1.93_{-0.07}^{+0.06}\,_{-0.10}^{+0.14}$
 & $10.63_{-0.17}^{+0.13}\,_{-0.37}^{+0.48}$ \\
$\sigma_{\rm N^3LL+NLO}$ & $1.96_{-0.05}^{+0.05}\,_{-0.11}^{+0.14}$
 & $10.81_{-0.08}^{+0.10}\,_{-0.37}^{+0.48}$ \\
\hline
 & LHC (14\,TeV, SUSY point $P_1$) & LHC (14\,TeV, SUSY point $P_2$) \\
\hline
$\sigma_{\rm LO}$ & $28.14_{-0.34}^{+0.25}\,_{-0.94}^{+0.70}$
 & $1.88_{-0.08}^{+0.09}\,_{-0.08}^{+0.07}$ \\
$\sigma_{\rm NLL}$ & $33.36_{-3.45}^{+4.60}\,_{-1.01}^{+1.10}$
 & $2.24_{-0.20}^{+0.27}\,_{-0.08}^{+0.09}$ \\
$\sigma_{\rm NLO}$ & $36.65_{-0.35}^{+0.45}\,_{-1.19}^{+1.28}$
 & $2.45_{-0.05}^{+0.05}\,_{-0.10}^{+0.11}$ \\
$\sigma_{\rm NNLL+NLO}$ & $37.16_{-0.46}^{+0.36}\,_{-1.03}^{+1.30}$
 & $2.47_{-0.03}^{+0.03}\,_{-0.08}^{+0.11}$ \\
$\sigma_{\rm N^3LL+NLO}$ & $37.80_{-0.12}^{+0.25}\,_{-1.05}^{+1.32}$
 & $2.51_{-0.02}^{+0.02}\,_{-0.08}^{+0.11}$ \\
\hline
\end{tabular}
\vspace{2mm}
\caption{\label{tab2}
Total cross sections in fb. The first error refers to the perturbative uncertainties associated with scale variations, the second to PDF uncertainties.}
\end{table}

The results shown in the first three blocks of Table~\ref{tab2} refer to the production of a slepton $\tilde l_L$ with mass $m_{\tilde l_L}=180$\,GeV (SUSY parameter point $P_1$, with $m_{\tilde q}=600$\,GeV and $m_{\tilde g}=750$\,GeV) at the Tevatron and the LHC. Note the relevance of the NLO correction, which amounts to around 40\% for the Tevatron and 30\% for the LHC. As expected, the resummation effects are larger at the Tevatron, where they amount to a 7\% enhancement of the NNLL+NLO cross section compared with the NLO result. At the LHC the resummation gives a smaller 3\% additional contribution to the total cross section. The additional contribution of the N$^3$LL+NLO result compared to the NNLL+NLO approximation is small, below 1\%, but performing the resummation at N$^3$LL order helps to further reduce the scale uncertainty.

\begin{figure}[t]
\begin{center}
\includegraphics[width=0.48\textwidth]{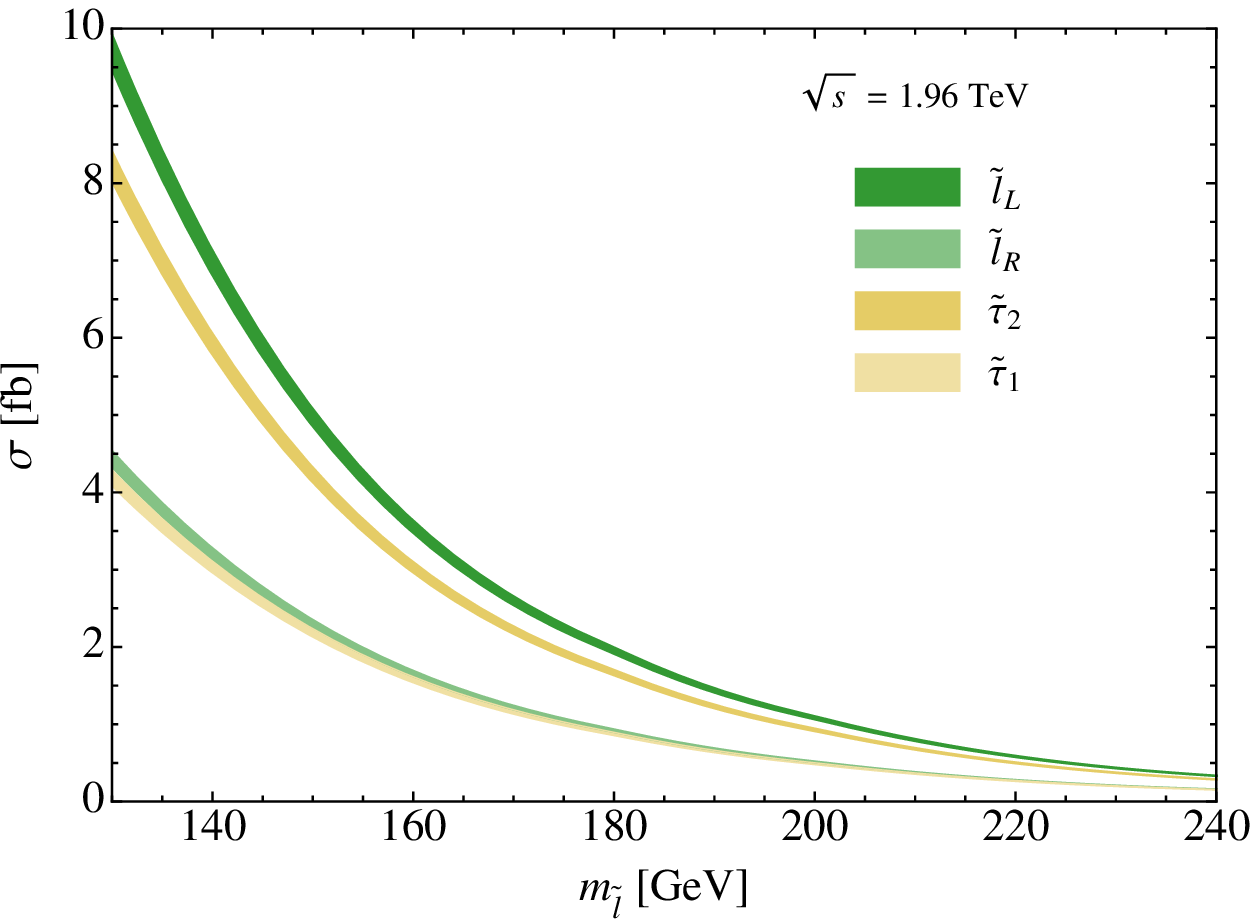} \\[4mm]
\includegraphics[width=0.48\textwidth]{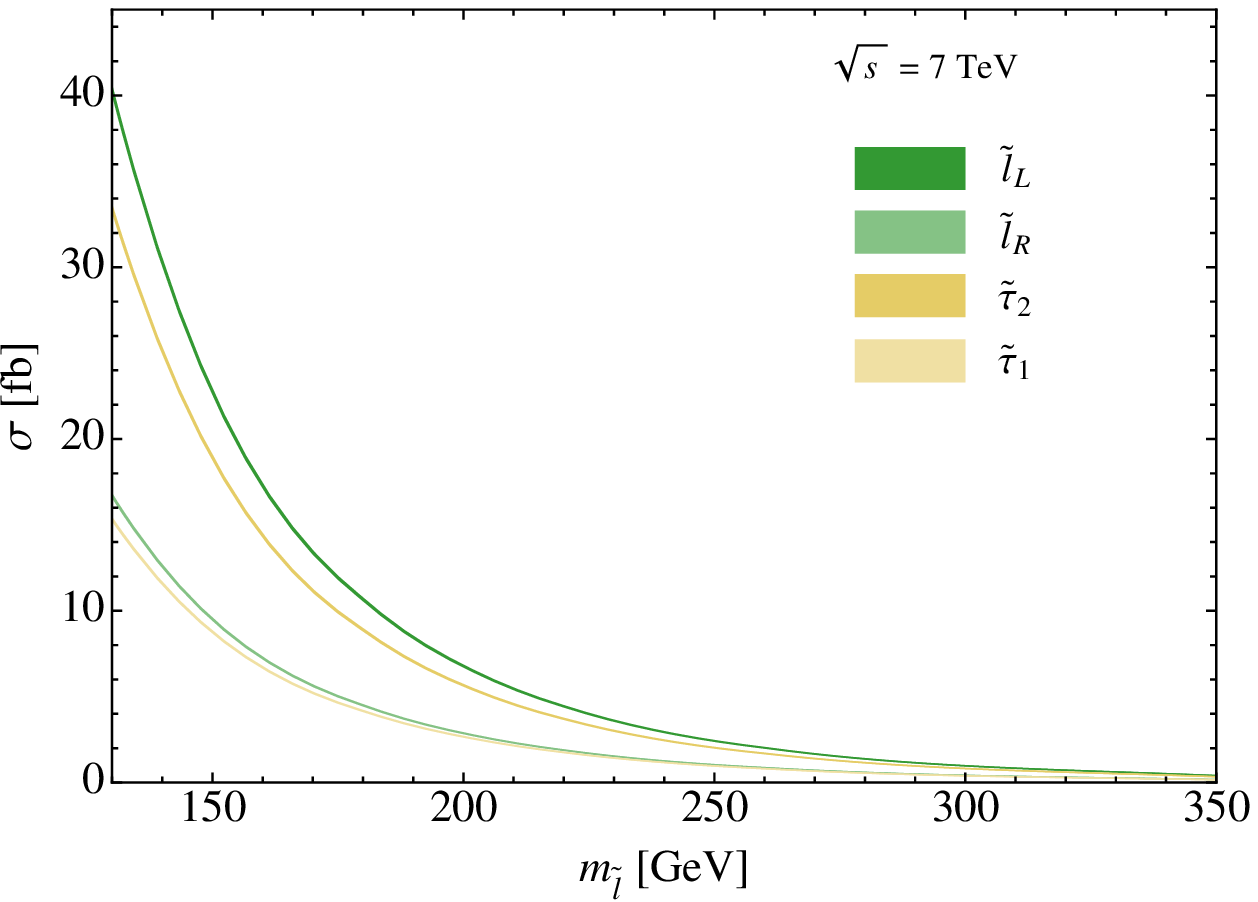}
\quad
\includegraphics[width=0.48\textwidth]{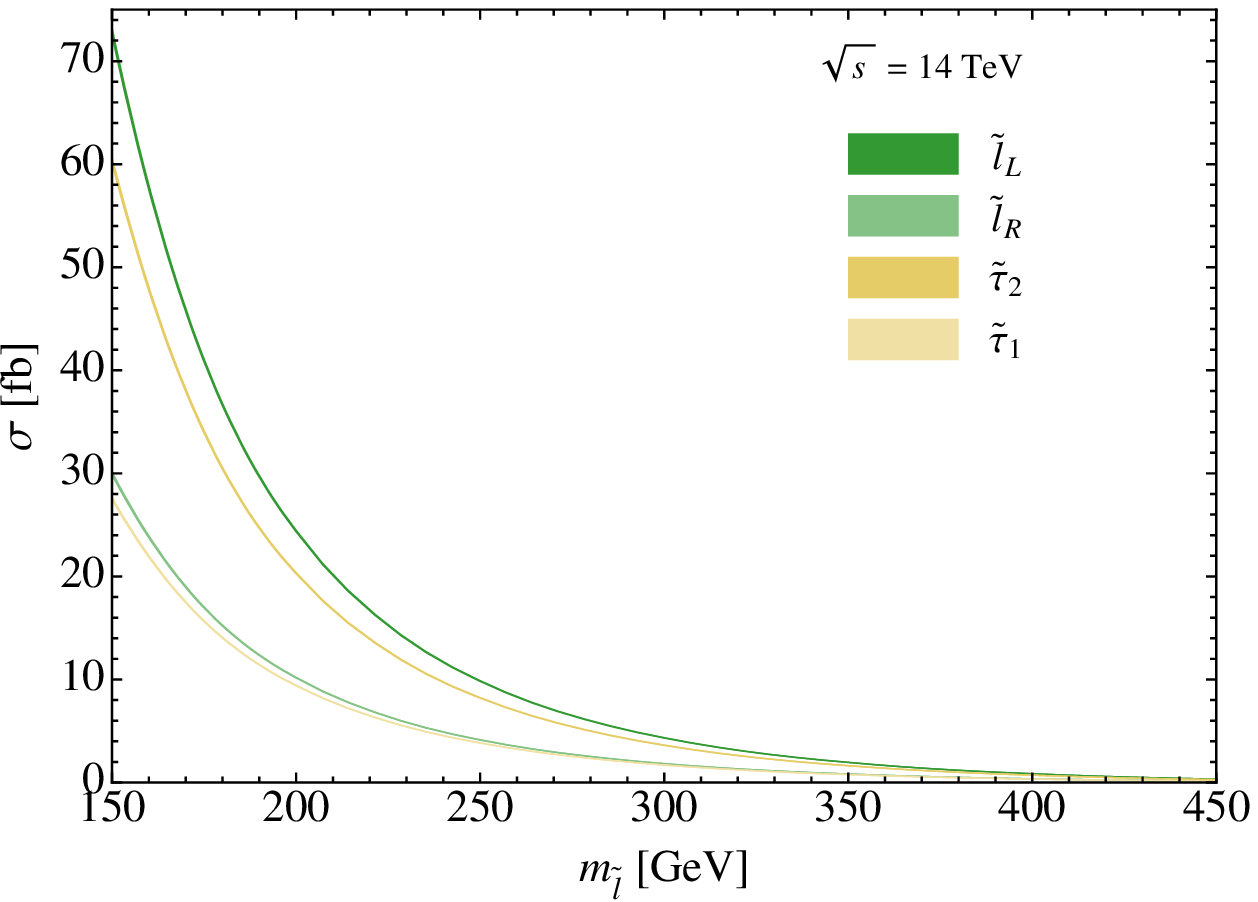}
\end{center}
\vspace{-4mm}
\caption{\label{fig10}
Total cross sections for slepton-pair production as a function of the slepton masses.}
\end{figure}

Since the effect of soft-gluon resummation becomes more important for higher invariant masses, in Table~\ref{tab2} we provide also the total cross section for a heavier slepton $\tilde l_L$ with mass $m_{\tilde l_L}=360$\,GeV (SUSY parameter point $P_2$, with $m_{\tilde q}=1200$\,GeV and $m_{\tilde g}=500$\,GeV). We only show results for the LHC with $\sqrt{s}=14$\,TeV, because given the small cross sections it would not be possible to observe the production of such heavy sleptons at the Tevatron or during the low-energy phase of the LHC.

In Figure~\ref{fig10}, we show the matched N$^3$LL+NLO total cross section as a function of the slepton mass. We now consider different types of sleptons, $\tilde l_{L,R}$ (with $l=e,\mu$) and $\tilde\tau_{1,2}$. For the staus, we assume a mixing angle $\theta_{\tilde\tau}=70^\circ$. The cross sections fall off steeply with the slepton masses. At the Tevatron it will be difficult to observe sleptons with masses exceeding about 250\,GeV, while at the LHC it should be possible to observe slepton-pair production up to masses in the range 300--400\,GeV. In the upper plot we show the cross sections at the Tevatron, focusing on the low mass region. The lower plots refer to the LHC at 7 and 14\,TeV of center-of-mass energy. At the same value of slepton mass, the slepton $\tilde l_L$ has the larger cross section, while the slepton $\tilde l_R$ has the lowest cross section. The cross sections for the production of staus lie in between. We observe that owing to the small scale uncertainty at N$^3$LL+NLO, it would be straightforward to extract the masses of the sleptons from measurements of the corresponding total cross sections.

We have compared our predictions for the LO and NLO fixed-order total cross sections for slepton production with results provided by the program {\tt PROSPINO} \cite{Beenakker:1999xh}, finding agreement. Despite the fact that it is difficult to compare our resummed results with those presented in \cite{Bozzi:2007qr}, because contrary to these authors we do not consider squark mixing, we still find a reasonable agreement. We emphasize that the method developed here allows us to resum soft gluons up to the N$^3$LL order, so that we are able to get a smaller scale uncertainty compared with \cite{Bozzi:2007qr}. 

The main result of our analysis is that, using soft-gluon resummation techniques, we can reduce the theoretical uncertainty related to scale variations below the percent level, making it a subdominant source of error.  This is evident from the results collected in Table~\ref{tab2}, which show that the error due to uncertainties in the PDFs becomes dominant beyond NLO. At this level of precision, one may ask whether the NNLO subleading terms could also become relevant. While the full NNLO corrections have not yet been calculated for the case of slepton pair production, we can estimate their relevance by considering the case of Drell-Yan production of lepton pairs, for which the NNLO corrections are availble \cite{Anastasiou:2003yy}. In this case, we find that the additional correction due to the NNLO subleading terms amounts to at most 1\% of the NLO cross section. It is thus of the same order as the scale uncertainty we find at N$^3$LL+NLO.

\section{Conclusions}
\label{concl}

If SUSY exists, the pair production of sleptons is an interesting discovery process at hadron colliders. Given the simple signature, which for a wide range in SUSY parameter space would consist of a pair of energetic leptons plus missing energy, it should be relatively easy to detect this process despite its cross section being an order of magnitude smaller than corresponding QCD processes, such as squark and gluino pair production. We have analyzed the slepton-pair production cross sections at the Tevatron and the LHC, along with the related cross section for the Drell-Yan production of a lepton pair, with the aim of obtaining accurate predictions by taking into account the effects of soft-gluon resummation. This was done by using methods of effective field theory, which allow us to perform the resummation directly in momentum space. The factorized cross sections in the partonic threshold region are expressed in terms of Wilson coefficiens of SCET operators. Solving the RG equations obeyed by these operators allows us to resum the large logarithms arising due to soft gluon emissions to all orders in the strong coupling constant.

We have extended the results available in literature in various directions. For the Drell-Yan process, we have calculated the effect of virtual SUSY QCD corrections at one-loop order. In the case of  slepton-pair production, we have extended previous results by performing the resummation up to the N$^3$LL level. Moreover, given the fact that we perform the resummation not in Mellin moment space but directly in momentum space, our results constitute an independent estimation of soft-gluon effects. We have provided a detailed phenomenological analysis, presenting results valid for the Tevatron and the LHC with $\sqrt{s}=7$ and 14\,TeV. We find that the SUSY QCD corrections due to the exchange of squark and gluinos are very small in comparison with the SM QCD corrections, and also with the uncertainty in the perturbative calculations. It would therefore be challenging to observe the effects of virtual SUSY particles in the Drell-Yan rapidity and invariant mass distributions.

We find that soft-gluon resummation has a small effect on the total cross sections for slepton-pair production, ranging from 7\% at the Tevatron to 3\% at the LHC for a slepton mass of 180\,GeV. This is a consequence of the fact that resummation effects are important only for large values of the invariant mass, corresponding to a region where the invariant-mass distribution is very small and gives a tiny contribution to the total cross section. Resummation is therefore more important for colliders where $\tau=M^2/s$ is larger, i.e.\ the Tevatron, or for higher slepton masses. On the other hand, it still proves useful for the reduction of the theoretical uncertainty due to scale variations. We find that the scale uncertainty reduced by about a factor of 2 when going from the fixed-order NLO calculation to the N$^3$LL+NLO result. The dominant uncertainties then arise from other sources, such as the imperfect knowledge of the PDFs, the parameters of the SUSY spectrum, Monte-Carlo modeling of the experimental acceptances for these SUSY final states, etc. Therefore, the improvement given by soft-gluon resummation constitutes only a first step towards a better derivation of the slepton masses from measurements of the total cross section.

\subsubsection*{Acknowledgements}

We are grateful to V.~Ahrens, A.~Ferroglia, B.~Pecjak, and L.L.~Yang and for useful comments and
suggestions. This work was supported in part by the DFG Graduate Training Center GRK~1581. L.V.\
acknowledges the Alexander-von-Humboldt Foundation 
for support.

\newpage
\appendix
Here we provide explicit expressions for the loop functions $f_B$ and $f_C$ appearing in (\ref{a16}), distinguishing two kinematical regimes. Below the squark production threshold, i.e.\ for $M^2\le 4m_{\tilde q}^2$, the two functions are real, while above threshold they develop an imaginary part. We first give results for the function $f_B$. Denoting $x=4m_{\tilde q}^2/M^2$, we obtain
\bea
   f_B(M^2,m_{\tilde q}^2) &=& 2\sqrt{x-1}\,\arctan\frac{1}{\sqrt{x-1}} \,; \quad
    x\ge 1 \,, \nn\\
   f_B(M^2,m_{\tilde q}^2) 
   &=& \sqrt{1-x} \left( \ln\frac{1+\sqrt{1-x}}{1-\sqrt{1-x}} - i\pi \right) ; \quad
    x<1 \,.
\eea
To express the function $f_C$ in a compact form, it is convenient to define 
\be
   y_0 = \frac{m_{\tilde q}^2-m_{\tilde g}^2}{M^2} \,, \qquad
   y_1 = \frac{m_{\tilde g}^2}{m_{\tilde g}^2-m_{\tilde q}^2} \,, \qquad
   y_\pm = \frac{1\pm\sqrt{1-x}}{2} \,.
\ee
For $M^2\le 4m_{\tilde q}^2$ we then obtain
\bea
   f_C(M^2,m^2_{\tilde q},m^2_{\tilde g}) 
   &=& \mathrm{Li}_2\bigg(\frac{y_0-1}{y_0-y_1}\bigg)
    - \mathrm{Li}_2\bigg(\frac{y_0}{y_0-y_1}\bigg)
    + \mathrm{Li}_2\bigg(\frac{y_0}{y_0-y_+}\bigg) \nn\\
   &&\mbox{}- \mathrm{Li}_2\bigg(\frac{y_0-1}{y_0-y_+}\bigg)
    + \mathrm{Li}_2\bigg(\frac{y_0}{y_0-y_-}\bigg)
    - \mathrm{Li}_2\bigg(\frac{y_0-1}{y_0-y_-}\bigg) \,,
\eea
while for $M^2>4m_{\tilde q}^2$ we get
\bea
   f_C(M^2,m^2_{\tilde q},m^2_{\tilde g}) 
   &=& \frac{\pi^{2}}{3} + \mathrm{Li}_2\bigg(\frac{y_0-1}{y_0-y_1}\bigg)
    - \mathrm{Li}_2\bigg(\frac{y_0}{y_0-y_1}\bigg) \nn\\
   &&\hspace{-2.0cm}
    \mbox{}+ \mathrm{Li}_2\bigg(\frac{y_0}{y_0-y_+}\bigg)
    + \mathrm{Li}_2\bigg(\frac{y_0-y_+}{y_0-1}\bigg)
    + \frac12\,\bigg[ \ln\bigg(\frac{y_0-1}{y_0-y_+}\bigg) + i\pi \bigg]^2 \nn\\
   &&\hspace{-2.0cm}
    \mbox{}+ \mathrm{Li}_2\bigg(\frac{y_0}{y_0-y_-}\bigg)
    + \mathrm{Li}_2\bigg(\frac{y_0-y_-}{y_0-1}\bigg)
    + \frac12\,\bigg[ \ln\bigg(\frac{y_0-1}{y_0-y_-}\bigg) - i\pi \bigg]^2 \,.
\eea
In the special case of equal masses, $m_{\tilde g}=m_{\tilde q}$, these results simplify significantly. We then obtain
\be
   c^{(1)}_{V,\rm SUSY}
   = C_F \left[ 3 - f_B(M^2,m_{\tilde q}^2)
    + \frac{2m_{\tilde q}^2}{M^2}\,f_C(M^2,m_{\tilde q}^2,m_{\tilde q}^2) \right] ,
\ee
where (with $x=4m_{\tilde q}^2/M^2$ as before)
\bea
   f_C(M^2,m_{\tilde q}^2,m_{\tilde q}^2) &=& - 2\arctan^2\frac{1}{\sqrt{x-1}} \,; \quad
    x\ge 1 \,, \nn\\
   f_C(M^2,m_{\tilde q}^2,m_{\tilde q}^2)
   &=& \frac12 \left( \ln\frac{1+\sqrt{1-x}}{1-\sqrt{1-x}} - i\pi \right)^2 ; \quad
    x<1 \,.
\eea

\bibliography{paperbib290411}
\bibliographystyle{JHEP-2}

\end{document}